%% file: main.tex
\documentclass{article}

\usepackage[accepted]{icml2024}

\usepackage[english]{babel}
\usepackage{amsmath}
\usepackage{bm}
\usepackage{amssymb}
\usepackage{graphicx}
\usepackage{url}
\usepackage{hyperref}
\usepackage[capitalize,noabbrev]{cleveref}
\usepackage{booktabs}
\usepackage{wrapfig}
\usepackage[nolist]{acronym}
\usepackage{todonotes}
\usepackage{tcolorbox}
\usepackage{multirow}
\usepackage{siunitx}
\usepackage{markdown}

\newcounter{colorToggle}

\newcommand{\add}[1]{
    \ifodd\value{colorToggle}
        \textcolor{blue}{#1}
    \else
        #1
    \fi
}

\newif\ifshowdelete

\showdeletefalse 

\newcommand{\delete}[1]{\ifshowdelete\textcolor{red}{#1}\fi}

\usepackage{microtype}
\usepackage{graphicx}
\usepackage{subfigure}
\usepackage{booktabs} 

\usepackage{hyperref}

\input{math_commands.tex}

\newif\ifcomments
\commentstrue  

\usepackage{amsmath}
\usepackage{amssymb}
\usepackage{mathtools}
\usepackage{amsthm}

\usepackage[capitalize,noabbrev]{cleveref}

\begin{acronym}
  \acro{SI}{Stealthy Imitation}
  \acro{RL}{reinforcement learning}
  \acro{BC}{behavioral cloning}
  \acro{KD}{knowledge distillation}
  \acro{SAC}{soft actor-critic}
  \acro{MLP}{multi-layer perceptron}
  \acro{KL}{Kullback-Leibler}
  \acro{DFME}{data-free model extraction}
  \acro{KDE}{kernel density estimation}
  \acro{GAIL}{generative adversarial imitation learning}
  \acro{TQC}{truncated quantile critics}
\end{acronym}

\theoremstyle{plain}

\theoremstyle{definition}

\theoremstyle{remark}

\icmltitlerunning{Stealthy Imitation: Reward-guided Environment-free Policy Stealing}

\begin{document}

\twocolumn[
\icmltitle{\raggedright Stealthy Imitation: Reward-guided Environment-free Policy Stealing}



\icmlsetsymbol{equal}{*}

\begin{icmlauthorlist}
\icmlauthor{Zhixiong Zhuang}{yyy,comp}
\icmlauthor{Maria-Irina Nicolae}{comp}
\icmlauthor{Mario Fritz}{sch}
\end{icmlauthorlist}

\icmlaffiliation{yyy}{Graduate School of Computer Science, Saarland University, Saarbrücken, Germany}
\icmlaffiliation{comp}{Bosch Center for Artificial Intelligence, Robert Bosch GmbH, Renningen, Germany}
\icmlaffiliation{sch}{CISPA Helmholtz Center for Information Security, Saarbrücken, Germany}

\icmlcorrespondingauthor{Zhixiong Zhuang}{zhixiong.zhuang@bosch.com}

\icmlkeywords{Machine Learning, ICML}

\vskip 0.3in
]



\printAffiliationsAndNotice{}

\begin{abstract}
\input{sections/00_Abstract/abstract}
\end{abstract}

\section{Introduction}
\label{sec:introduction}
\input{sections/01_Introduction/introduction}

\section{Related Work}
\label{sec:relatedworks}
\input{sections/02_RelatedWorks/relatedworks}

\section{Threat Model}
\label{sec:problemstatement}
\input{sections/03_ProblemStatement/problemstatement}

\section{Approach: Stealthy Imitation}
\label{sec:approach}
\input{sections/05_Approach/approach}

\section{Experiments}
\label{sec:experiments}
\input{sections/06_Experiment/00_experiments}

\section{Discussion}
\label{sec:discussion}

\input{sections/07_Discussion/discussion}

\section{Conclusion}
\input{sections/08_Conclusions/conclusions}


\section*{Acknowledgements}
We acknowledge the support and funding by Bosch AIShield. This work was also partially funded by ELSA – European Lighthouse on Secure and Safe AI funded by the European Union under grant agreement No. 101070617, as well as the German Federal Ministry of Education and Research (BMBF) under the grant AIgenCY (16KIS2012).

\section*{Impact Statement}
This paper presents work aimed at raising awareness among developers and those involved in deployment about the risks associated with Machine Learning. While we are demonstrating an attack in this work, we use a hypothetical simulation setup that enables the evaluation of generic attack vectors that do not target particular systems and do not cause harm. Therefore, no responsible disclosure procedure is necessary or even applies to our work. However, given the importance of the problem and the deployment of \ac{RL} in control systems, we believe it is now the time to evaluate such attacks and inform developers about potential risks. The potential negative societal impacts include the possibility that the proposed attack could be used against real control systems. However, we also propose a defense and encourage policy owners to use it. Similar to commonly published attacks in ML and security and privacy venues, our goal is to provide a novel evaluation that has the goal of, among others, improving the safety of such systems. The authors strictly comply with the ICML Code of Conduct\footnote{https://icml.cc/public/CodeOfConduct}.



\bibliography{my_bib}
\bibliographystyle{icml2024}

\newpage
\appendix
\onecolumn

\section{Algorithms}
\label{app:alg}
\input{sections/Appendix/01_algorithm}

\section{Environment and Victim Policy}
\label{app:envandvic}
\input{sections/Appendix/02_EnvAndVictim}

\section{Compute Resources}
\label{app:compute}
\input{sections/Appendix/15_compute}

\section{Influence of Model Architecture}
\label{app:advarch}
\input{sections/Appendix/03_ChoiceArch}

\section{Variability of Stealthy Imitation}
\label{app:variability_si}
\input{sections/Appendix/05_Variability_of_Stealthy_Imitation}

\section{Performance of the Reward Discriminator}
\label{app:reward_performance}
\input{sections/Appendix/09_reward_model_loss}

\section{Adaptation of DFME}
\label{app:adaptation_of_dfme}
\input{sections/Appendix/14_DFME}

\section{Robustness to Distribution Approximation Errors}
\label{app:center_scale}
\input{sections/Appendix/04_varying_center_scale}



\section{Risk of Exposing Distribution}
\label{app:risk_of_exposing_dist}
\input{sections/Appendix/13_risk_of_exposing_dist}

\section{Discussion on Defense}
\label{app:discussion_on_defense}
\input{sections/Appendix/11_discussion_on_defense}


\section{Impact of Probabilistic State Distribution Model}
\label{app:full_gaussian_distribution}
\input{sections/Appendix/18_full_gaussian_distribution}

\newpage
\section{Underlying Distribution}
\label{app:underlying_dist}
\input{sections/Appendix/12_underlying_distribution}


\end{document}

%% file: math_commands.tex
\def\eqref#1{equation~\ref{#1}}

\def\1{\bm{1}}

\def\vmu{{\bm{\mu}}}
\def\vsigma{{\bm{\sigma}}}

\def\va{{\bm{a}}}

\def\vs{{\bm{s}}}

\def\vz{{\bm{z}}}

\def\mSigma{{\bm{\Sigma}}}

\DeclareMathAlphabet{\mathsfit}{\encodingdefault}{\sfdefault}{m}{sl}
\SetMathAlphabet{\mathsfit}{bold}{\encodingdefault}{\sfdefault}{bx}{n}

\def\gD{{\mathcal{D}}}

\def\gN{{\mathcal{N}}}

\def\gR{{\mathcal{R}}}

\def\sR{{\mathbb{R}}}
\def\sS{{\mathbb{S}}}

\newcommand{\E}{\mathbb{E}}
\newcommand{\Ls}{\mathcal{L}}

\newcommand{\KL}{D_{\mathrm{KL}}}

\DeclareMathOperator*{\argmin}{arg\,min}

%% file: sections/00_Abstract/abstract.tex
Deep reinforcement learning policies, which are integral to modern control systems, represent valuable intellectual property. The development of these policies demands considerable resources, such as domain expertise, simulation fidelity, and real-world validation. These policies are potentially vulnerable to model stealing attacks, which aim to replicate their functionality using only black-box access.
In this paper, we propose Stealthy Imitation, the first attack designed to steal policies without access to the environment or knowledge of the input range. This setup has not been considered by previous model stealing methods.
Lacking access to the victim's input states distribution, Stealthy Imitation fits a reward model that allows to approximate it.
We show that the victim policy is harder to imitate when the distribution of the attack queries matches that of the victim.
We evaluate our approach across diverse, high-dimensional control tasks and consistently outperform prior data-free approaches adapted for policy stealing.
Lastly, we propose a countermeasure that significantly diminishes the effectiveness of the attack.\footnote{The project page is at \href{https://zhixiongzh.github.io/stealthy-imitation/}{https://zhixiongzh.github.io/stealthy-imitation}.}




%% file: sections/01_Introduction/introduction.tex
Neural networks trained with \acf{RL}, known as deep \ac{RL} policies, are increasingly employed in control systems due to the exceptional performance and automation capabilities. Examples include DeepMind’s use of \ac{RL} in cooling control systems~\cite{luo2022controlling}, throttle valve control in combustion engines~\cite{bischoff2013learning}, and Festo Robotino XT robot control~\cite{bischoff2014policy}. Developing a reliable deep \ac{RL} policy requires substantial resources, including expertise in training, precise simulation, and real-world testing; the resulting policy becomes important intellectual property. However, neural network models are vulnerable to stealing attacks~\citep{tramer2016stealing,knockoff,dfme} that attempt to copy the functionality of the model via black-box query access. The risks posed by such attacks \add{in control systems} are multifaceted, including unauthorized model usage, exposure of sensitive information, and further attacks that can lead to denial of service, operational failures, or even physical damage on the equipment or surroundings.

\begin{figure}[t]
    \centering
    \includegraphics[width=0.45\textwidth]{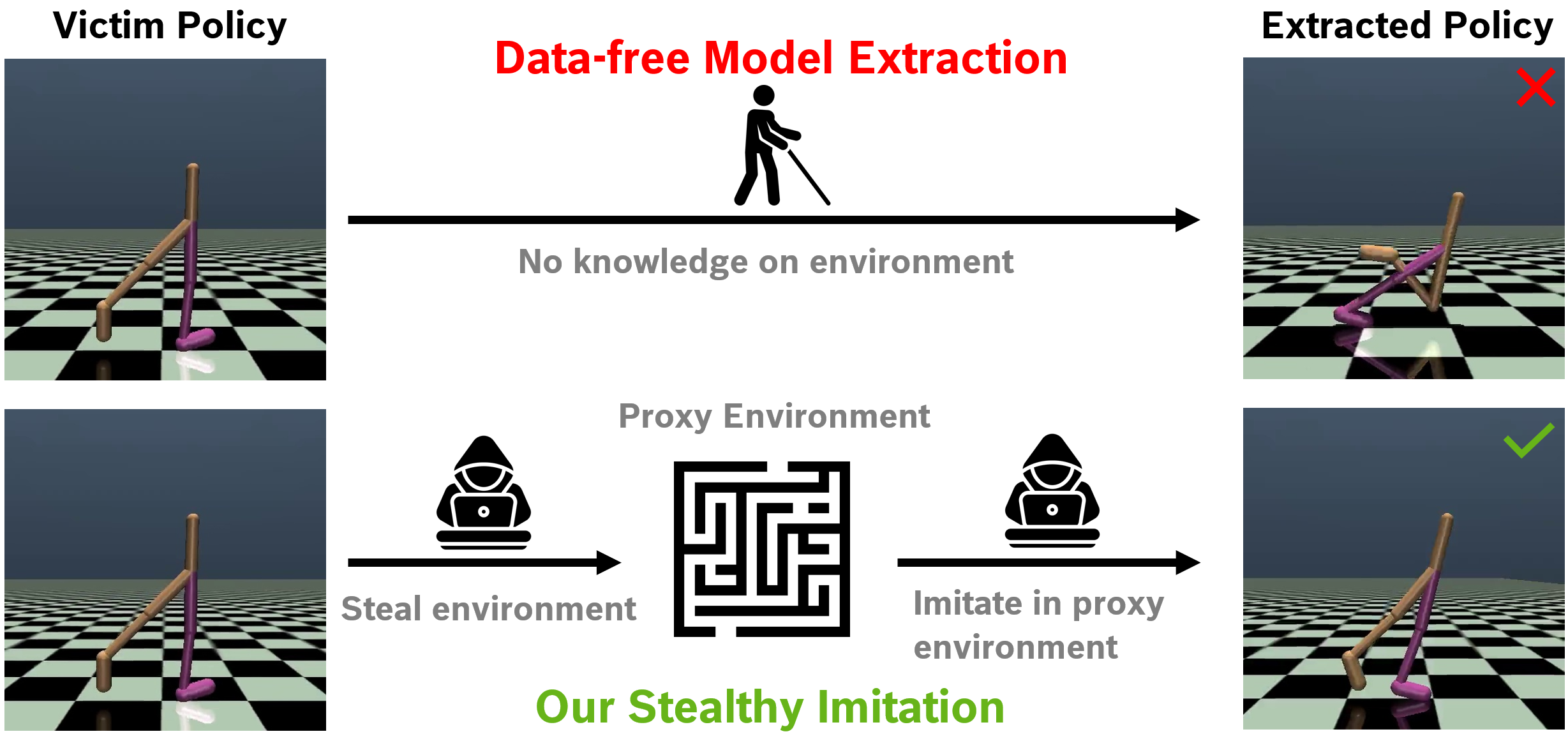}
    \caption{Traditional data-free model extraction fails in control systems due to the unknown environment with varying sensors. \acl{SI} effectively extracts policies by stealing the environment first.}
    \label{fig:illustraction}
\end{figure}

Model theft typically consists of two steps. First, a transfer dataset is created by querying the victim model with publicly available data~\citep{knockoff}, random noise~\citep{tramer2016stealing}, or samples synthesized by a neural network~\citep{dfme}, and recording the model predictions as pseudo-labels. The latter two methods fall under the category of data-free model stealing. After this querying phase, the attackers train their own model via supervised learning, treating the pseudo-labels as ground truth for their samples.

Control systems, such as industrial automation, remotely controlled drones or robots, pose additional challenges for model stealing. A policy perceives states and rewards (also known as the environment), based on which it decides the next action to take. In this context, the attacker can potentially send queries to the system, but does not have access to the environment.
Data-free stealing attacks hold the promise of environment-free policy stealing. While existing data-free attacks have proven effective in the image domain, they operate under the assumption that the attacker knows the valid input range. For instance, valid image pixels are assumed to be in the range of $[0,255]$.
However, such prior knowledge is difficult to acquire in control systems or other applications, due to the distinct semantics and scales of components within the measured state.
As a consequence, policy stealing becomes more difficult.

To address this challenge, 
we introduce \ac{SI}, the first environment-free policy stealing attack. Our method solves the two fundamental difficulties of this task: (i) the necessity of accurately estimating the input range and distribution of the states visited by the victim policy, and (ii) the identification of a metric that allows the attacker to evaluate the estimated distribution, and thus its own performance in stealing the policy.
These advancements collectively enable a more robust and efficient policy stealing attack. Notably, the derived distribution remains applicable even when the victim updates their policy without altering the training distribution, offering potential savings in query budget for subsequent attacks. Furthermore, it enables the attacker to access confidential information, like sensor types or preprocessing methods. The information could potentially aid in the development of their own control system. The superior performance of our method over traditional data-free model extraction is shown in \cref{fig:illustraction}.


\noindent\textbf{Contributions.}
(i) We introduce a more general and realistic threat model adapted to control systems, where the attacker lacks access to the environment and to the valid input ranges of the policy. (ii) We propose \acf{SI}, the first reward-guided environment-free policy stealing method under minimal assumptions.
We show our attack to be effective on multiple control tasks.
(iii) We introduce the first proxy metric to measure the quality of the estimated distribution. We empirically and statistically validate its correlation with the divergence between the estimated distribution and the actual state distribution of the victim policy.
(iv) We develop a defense that is able to counter the proposed attack, thus offering a practical solution for practitioners.


%% file: sections/02_RelatedWorks/relatedworks.tex
\noindent\textbf{Knowledge distillation.}
\Ac{KD} was initially designed for model compression, aiming to approximate a large neural network (commonly referred to as the teacher model) with a more compact model (the student model). This facilitates deployment on hardware with limited computational capabilities~\citep{dodeep?,hintonkd}. Unlike our work, which adopts an adversarial view, \ac{KD} typically presumes access to the teacher model's original training dataset, enabling the student model to learn under the same data distribution. When the dataset is large or sensitive, some methods opt for surrogate datasets~\citep{datafreekd}. Others eliminate the need for it by employing data generators in data-free \ac{KD} approaches~\citep{dfad, zeroshotkd}. These methods often assume white-box access to the teacher model for backpropagation, which is a major difference with our setup.

\noindent\textbf{Model stealing.}
Model stealing focuses on adversarial techniques for the black-box extraction of a victim model (equivalent to the teacher model in \ac{KD})~\citep{tramer2016stealing, knockoff}. The attacker, who aims to create a surrogate model (analogous to the student in \ac{KD}), lacks access to the original training dataset of the victim model. Most existing methods explore data-free stealing, drawing inspiration from data-free \acl{KD}, but lacking the means to use the victim model to train a data generator. These techniques estimate the gradient of the victim model for training their generator and encourage query exploration by synthesizing samples that maximize the disagreement between victim and attacker model~\citep{dfme_hardlabel, dualstudent,dfme}. While much work has been conducted in image-based domains, limited research exists on model stealing in the context of reinforcement learning~\citep{behzadan2019adversarial,rlstealingforfun}. Our approach sidesteps the need for environment access and specific knowledge of the \ac{RL} algorithm employed by the victim. Existing defenses primarily focus on detecting stealing attacks~\citep{juuti2019prada,kesarwani2018model} or perturbing model predictions~\citep{tramer2016stealing,Orekondy2019PredictionPT}. Our proposed defense falls in the latter category: the policy perturbs its outputs when the query falls outside the valid input range.

\noindent\textbf{Imitation learning.}
Imitation learning aims to train agents to emulate human or expert model behavior. Within this domain, there are two main methodologies. The first is \ac{BC}, which treats policy learning as a supervised learning problem, focusing on state-action pairs derived from expert trajectories~\citep{bc}. The second is inverse reinforcement learning, which seeks to discover a cost function that renders the expert's actions optimal~\citep{russell1998learning, ng2000algorithms}. Another method of interest is \ac{GAIL}, which utilizes adversarial training to match the imitating agent's policy to that of the expert. Notably, \ac{GAIL} achieves this alignment using collected data and does not need further access to the environment~\citep{gail}. Our work deviates from these imitation learning approaches, as we do not require access to the interaction data between the expert policy, i.e., the victim for us, and its environment.

%% file: sections/03_ProblemStatement/problemstatement.tex
In this section, we formalize the threat model for black-box policy stealing in the context of deep \ac{RL} policies used in control systems. First, we introduce preliminary concepts and notations. Then, we formalize the victim's policy. Finally, we outline the attacker's knowledge and the relevance of this threat model to real world attacks.

\noindent\textbf{Notations.} 
In the context of deep \ac{RL}, a policy or agent, is denoted by \( \pi \) with accepting state $\vs$, and predicting an action $\va$, such that $\va=\pi(\vs)$. A trajectory $ \tau \sim \pi $ consists of a sequence of states and actions collected from the interaction between policy and environment. We represent the initial state distribution as $ \rho_0 $, and the environment's state transition function as $f$, such that $\vs_{t+1} = f(\vs_t,\va_t)$. The return, or cumulative reward, for a trajectory is represented as $ R(\tau)$, while $S$ is the distribution of states visited by the deployed policy.

\noindent\textbf{Victim policy.} 
We consider a victim operating a deep \ac{RL} policy, \( \pi_v \), trained to optimize a particular control objective, accepting a state $\vs \in \sR^n$ and predicting the action $\va^* \in \sR^k$ within the range of $\left[ -1,1 \right]$ \add{at each time step}. \add{Note that victim policies accepting images as input are out of scope, since their input range is typically known (e.g., $[0, 255]$).}
The environment is fully observable by the victim policy.
The performance of the policy is quantified using the expected return \(\E_{\tau_v \sim \pi_v}[R(\tau_v)]\) in the deployed setting. \( S_v \) represents the distribution of states visited by \( \pi_v \).

\noindent\textbf{Goal and knowledge of the attacker.} 
We take on the role of the attacker, with the goal of training a surrogate policy $\pi_a$, predicting action $\hat{\va}$, to replicate the functionality of the victim policy $\pi_v$ for similar return in the environment. The attacker possesses black-box access to $\pi_v$ by querying states and obtaining actions as responses. The total amount of queries is represented as $B$. However, the attacker lacks knowledge on several key aspects: (i) the internal architecture and \ac{RL} training algorithm of $\pi_v$, (ii) the environment setup, including the initial state distribution $\rho_0$, the state-transition function $f$, and the reward function $R$, (iii) the semantics associated with the input and output spaces, (iv) the range of the inputs, as well as the state distribution $S_v$, and (v) the confidence score of all possible actions from the victim policy. This lack of knowledge makes policy stealing particularly challenging.

\noindent\textbf{Real-world relevance.}
\delete{Our threat model highlights the urgency of addressing vulnerabilities in deep \ac{RL} policy-based control systems, particularly those accessible remotely. In scenarios where attackers lack direct environment interaction, they resort to black-box policy queries to replicate system functionalities. 
The stolen policy can further be used to craft additional attacks against the system, resulting in denial of service, operational failures, or even physical damage to the equipment.
Moreover, input range estimation exposes sensitive information like sensor types or preprocessing methods, aiding attackers in developing their own control systems and creating potential for future security threats.}
\add{Our threat model highlights the urgency of addressing vulnerabilities in deep \ac{RL} policy-based control systems and targets two scenarios where attackers opt for an environment-free policy stealing method to exploit these systems. Firstly, the attacker might not know the environment when they access a networked victim policy they don’t own. Secondly, there are cases where interacting with the environment is possible but impractical and inefficient due to the risk of being detected, time, cost and damage concerns. With the widespread adoption of the Internet of Things (IoT) and control systems’ physical world impact, these scenarios are common. Once the policy is stolen, it can lead to additional system attacks, causing service denial, operational failures, or equipment damage. Moreover, successful state distribution estimation can reveal sensitive information like sensor types or preprocessing methods, helping attackers develop their own systems and posing future security threats. Therefore, it is important to investigate how attackers can steal policies without environment access and create defenses to safeguard control systems.}

%% file: sections/05_Approach/approach.tex
This section introduces the details of \acl{SI}.
The method overview in \Cref{sec:si_overview} is followed by an explanation of each of its components in \Cref{subsec:main_steps}.
\Cref{subsec:distribution_estimator} shows how to use the estimated distributions from prior steps to steal the target policy.
Lastly, we propose a defense that can make the attacker's goal more difficult to reach.

\subsection{Method Overview}
\label{sec:si_overview}

We introduce \acl{SI} as attacker that steals a policy without access to the environment or to the valid input range.
To achieve their goal, the attacker aims to optimize the surrogate policy $\pi_a$ to minimize the expected return difference between their own policy $\pi_a$ and that of the victim $\pi_v$ in the environment:

\begin{equation}
\label{eq:objective_on_traj}
\argmin_{\pi_a} \left|\underset{\tau_a \sim \pi_a}{\E}[R(\tau_a)] - \underset{\tau_v \sim \pi_v}{\E}[R(\tau_v)]\right|
\end{equation}

However, the attacker does not have access to the environment or the reward function. Instead, they can minimize the action difference between their policy and that of the victim on an estimated state distribution $S_a$ using a loss function $\Ls$ as a proxy for the reward:

\begin{equation}
\label{eq:objective_on_sa}
\argmin_{\pi_a} \underset{\vs \sim S_a}{\E}[\Ls(\pi_v(\vs),\pi_a(\vs))]
\end{equation}

The attacker's goal is thus to find both the victim policy and the appropriate distribution of states.
The \acl{SI} objective encourages exploration by maximizing the disagreement between the victim and attacker models:


\begin{equation}
\label{eq:numbering_objective}
\overset{\textbf{II}}{\arg\min_{\pi_a}\vphantom{\arg\max_{S_a}}} ~ \overset{\textbf{III \& IV}}{\arg\max_{S_a}\vphantom{\arg\min_{\pi_a}}}  \underset{\vs \sim S_a}{\E}\left[ \Ls(\pi_a(\vs), \overset{\textbf{I}}{\pi_v(\vs)}) \right].
\end{equation}


\begin{figure}[t]
    \centering
    \includegraphics[width=0.45\textwidth]{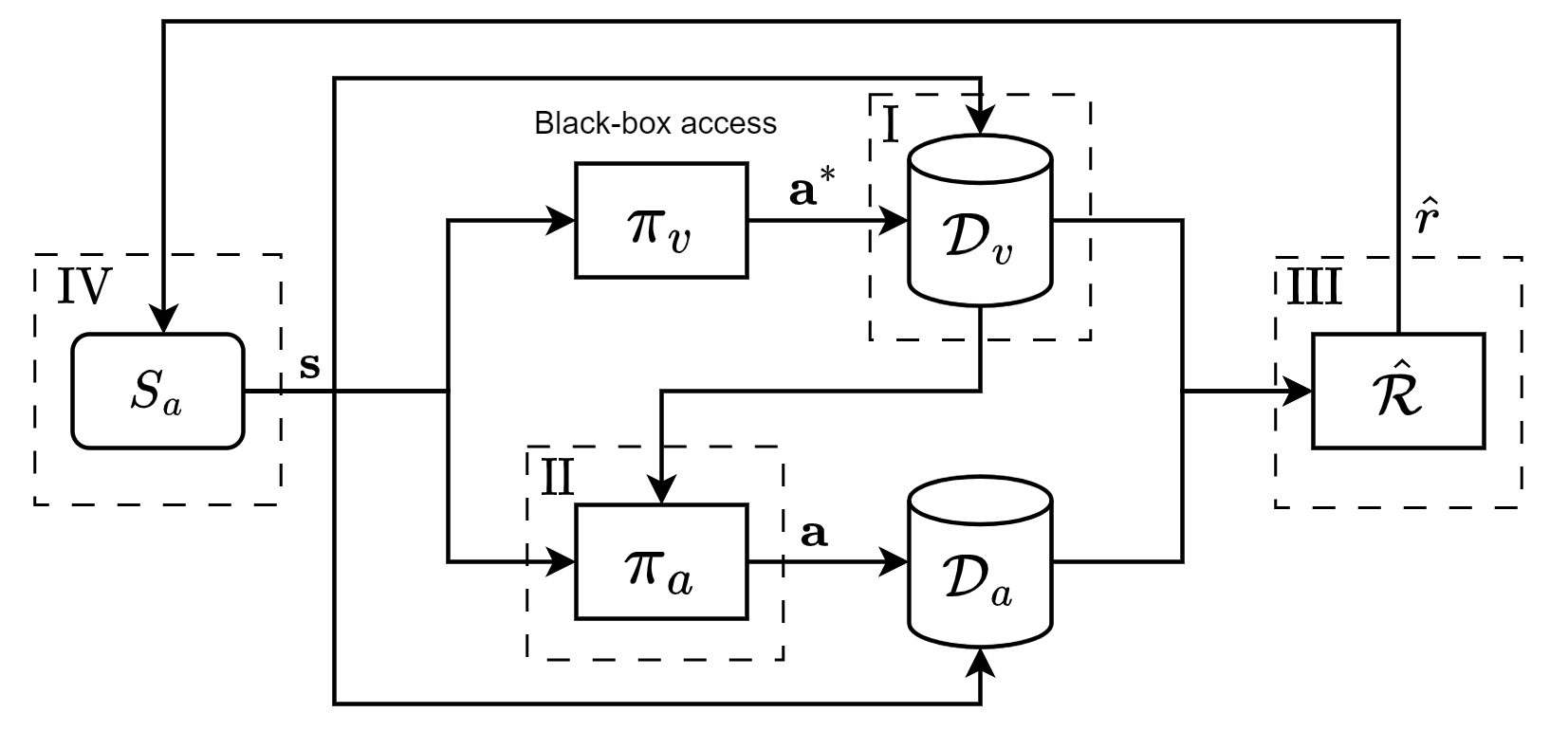}
    \caption{Overview of Stealthy Imitation that iteratively refines the estimated state distribution $S_a$.}
    \label{fig:SI_diagram}
\end{figure}

The core of Stealthy Imitation consists of four main steps repeated iteratively until the attacker query budget is consumed: (I) \textbf{transfer dataset construction} by querying the victim policy with states sampled from the estimated distribution $S_a$; (II) training the attacker policy $\pi_a$ via \textbf{behavioral cloning} to mimic the victim policy on the transfer dataset; (III) \textbf{reward model training} $\hat{R}$ to discriminate the behaviours of the victim and current attacker policy, and (IV) \textbf{reward-guided distribution refinement} to closer match the victim's state distribution using the proxy reward score on each query state.
Once the attacker's budget is exhausted, we train $\pi_a$ from scratch only on the best estimated distribution with the help of a distribution evaluator. The approach overview is depicted in~\Cref{fig:SI_diagram}. We detail each step in the following.

\subsection{State Distribution Estimation}
\label{subsec:main_steps}


\noindent\textbf{I. Transfer dataset construction.}
As the attacker has no knowledge of the state distribution of the victim $S_v$, we choose a \delete{multivariate normal} \add{diagonal Gaussian} distribution $\mathcal{N}(\vmu, \vsigma^2)$ \delete{with a diagonal covariance matrix} as estimate of the attacker distribution $S_a$ ($S_a(\vs; \vmu, \vsigma)$ from here on). States $\vs$ are sampled from this distribution and passed to the victim policy to obtain corresponding actions $\va^*$. The transfer dataset $\gD_v$ described below is split into training and validation for use in the subsequent method steps:

\begin{equation}
\label{eq:queryaction}
\begin{aligned}
\gD_v &= \{(\vs,\va^*)\}, \\
\text{where} \quad \vs &\sim S_a(\vs; \vmu,\vsigma),  \\
\text{and} \quad \va^* &= \pi_v(\vs).
\end{aligned}
\end{equation}

Lacking prior knowledge, $S_a$ is initialized with $\vmu = \mathbf{0}_n$ and $\vsigma = \mathbf{1}_n$. In each iteration, we use a dynamic query budget by multiplying a base budget $b_v$ with the average of $\vsigma$. This ensures sufficient learning in mimicking the actions of the victim policy, especially when the estimated $ \vsigma $ is large, thereby stabilizing the refinement process.

\noindent\textbf{II. Behavioral cloning.}
We follow the conventional step in model stealing to mimic the victim policy's behavior using the training split of the transfer dataset $\gD_v$. To this end, we employ behavioral cloning using Huber loss~\citep{huber_robust_1964}, which is robust to outliers like L1 and smooth and differentiable near the minimum like L2:

\begin{equation}
\label{eq:smoothl1}
\mathcal{L}_b(\hat{\va},\va^*) =
\begin{cases}
0.5(\hat{\va}-\va^*)^2, & \text{if } |\hat{\va}-\va^*| < 1 \\
|\hat{\va}-\va^*| - 0.5, & \text{otherwise}
\end{cases}
\end{equation}

\noindent\textbf{III. Reward model training.}
Our approach is driven by the intuition that the victim policy, while complex within its domain, behaves more simply outside it due to insufficient training. This simpler behavior can be more easily mimicked. Based on this intuition, we hypothesize that as the estimated state distribution $S_a$ approaches the victim's state distribution $S_v$, the complexity of the victim's responses increases. This makes it more difficult for the attacker policy $\pi_a$ to accurately imitate the victim. This hypothesis is supported empirically by the results in \Cref{subsec:analysis}.
To evaluate the difficulty of imitation from the state-action pairs from $\pi_a$ and $\pi_v$, we introduce a reward model $\hat{\gR}$, inspired by \ac{GAIL}~\citep{gail}. The role of $\hat{\gR}$ is to distinguish between state-action pairs generated by the victim and attacker policy. A more effective distinction suggests that the attacker's policy is more challenging to imitate accurately. To this end, we construct a dataset $\mathcal{D}_a $ using actions $\hat{\va}$ generated by $\pi_a(\mathbf{s}) $ after \ac{BC}, and train a reward model $\hat{\gR}$ by minimizing the loss function:

\begin{equation}
\begin{aligned}
\Ls_{r}(\vs,\va) = & \underset{(\vs, \hat{\va}) \sim \gD_a}{\E}[-\log(\hat{\gR}(\vs, \hat{\va}))] \\
& + \underset{(\vs, \va^*) \sim \gD_v}{\E}[-\log(1 - \hat{\gR}(\vs, \va^*))]. \label{eq:loss_reward}
\end{aligned}
\end{equation}


\noindent\textbf{IV. Reward-guided distribution refinement.}
We use the trained reward model from the previous step $\hat{\gR}$ to generate proxy reward values $ \hat{r}(\vs, \va^*) = -\log(\hat{\gR}(\vs, \va^*)) $ for each state-action pair.
A high reward value $ \hat{r}(\vs,\va^*) $ indicates that the attacker policy fails to effectively mimic the victim, suggesting that the state has higher probability in $S_v$. 
These reward values serve as weights for the corresponding samples $\vs$, which we use to recompute the parameters $\vmu'$ and $\vsigma'$ of the distribution for the next iteration, as follows:

\begin{equation}
\begin{aligned}
\label{eq:distrefine}
\vmu' &= \frac{\sum_{(\vs, \va^*) \in \gD_v} \hat{r}(\vs,\va^*) \cdot \vs}{\sum_{(\vs, \va^*) \in \gD_v} \hat{r}(\vs,\va^*)}, \\
\vsigma'^2 &= \frac{\sum_{(\vs, \va^*) \in \gD_v} \hat{r}(\vs,\va^*) \cdot (\vs-\vmu')^2}{\sum_{(\vs, \va^*) \in \gD_v} \hat{r}(\vs,\va^*)}.
\end{aligned}
\end{equation}

\subsection{Policy Stealing on the Estimated Distribution}
\label{subsec:distribution_estimator}

Since the attacker has no knowledge of the victim states' distribution $S_v$, we introduce a model $\pi_e$, which we term distribution evaluator.
This model helps assess the closeness between the attacker and victim distributions $ S_a $ and $ S_v $.
$\pi_e$ is trained via \acl{BC} and is reinitialized in each iteration to ensure its validation loss $\bar{\Ls}_b$ measures only the error of the current estimated distribution. Based on our hypothesis, a higher loss $ \bar{\Ls}_b $ is indicative of $ S_a $ closely mirroring $ S_v $. We only use $b_v$ samples of the transfer dataset \(\gD_v\) to train \(\pi_e\) instead of $b_v \times \bar{\vsigma}$. This ensures it is only affected by the distribution divergence without the influence of training data size. 
The total query budget $B$ comprises two parts: the budget used to refine the distribution and the reserved budget $B_r$ for training the final attacker policy on the best-estimated distribution. 
Once the first part of budget is exhausted, i.e., the algorithm is done iterating over steps I-IV, the parameters \( \tilde{\vmu} \) and \( \tilde{\vsigma} \) from the iteration that yielded the highest loss value are used to create an optimized transfer dataset using the reserved query budget $B_r$. Finally, \( \pi_a \) is subsequently retrained from scratch via \ac{BC} using this optimized dataset. \Cref{alg:SI} outlines the complete method; all the functions used are defined in \Cref{app:alg}.

\begin{algorithm}[tb]
\caption{\acl{SI}}
\label{alg:SI}
\begin{small}
\begin{algorithmic}[1]
   \STATE {\bfseries Input:} Victim policy \( \pi_v \) (blackbox access), total budget \( B \), reserved budget \( B_r \), base query budgets \( b_v \) and \( b_a \) for victim and attacker victims respectively in each iteration
   \STATE {\bfseries Output:} Trained attacker policy \( \pi_{a} \)
   \STATE Initialize attacker policy \( \pi_{a} \), distribution evaluator \( \pi_e \), reward model \( \hat{\gR} \), \( \vmu \gets \mathbf{0}_n \), \( \vsigma \gets \mathbf{I}_n \)
   \STATE Initialize proxy metric \( \tilde{\Ls} \gets -\infty \), consumed budget \( B_c \gets 0 \), and to be consumed budget \( b_c \gets b_v \)
   \REPEAT
      \STATE // I. Transfer dataset construction
      \STATE \( \gD_v \gets \text{QueryAction}(\pi_v, \vmu, \vsigma, b_c) \)
      \STATE // Evaluate current estimated distribution
      \STATE \( \bar{\Ls}_b \gets \text{DistributionEvaluate}(\gD_{v}, \pi_e, b_v) \)
      \STATE // Update parameters if current loss exceeds max loss
      \IF{\(  \bar{\Ls_b} > \tilde{\Ls} \)}
         \STATE \( \tilde{\gD},\tilde{\Ls},\tilde{\vmu},\tilde{\vsigma}  \gets \gD_v,  \bar{\Ls}_b, \vmu, \vsigma \)
      \ENDIF
      \STATE // II. Behavioral cloning
      \STATE \( \pi_{a} \gets \text{BehavioralCloning}(\gD_{v}, \pi_{a}, b_v \cdot \bar{\vsigma}) \)
      \STATE // III. Reward model training
      \STATE \( \gD_a \gets \text{QueryAction}(\pi_{a}, \vmu, \vsigma, b_a) \)
      \STATE \( \hat{\gR} \gets \text{TrainReward}(\gD_a, \gD_{v}, \hat{\gR}, b_v \cdot \bar{\vsigma}) \)
      \STATE // IV. Reward-guided distribution refinement
      \STATE \( \vmu, \vsigma \gets \text{DistRefine}(\gD_v, \hat{\gR}, b_v \cdot \bar{\vsigma}) \)
      \STATE \( B_c \gets B_c + b_c \)
      \STATE \( b_c \gets \max(b_v, b_v \cdot \bar{\vsigma}) \)
   \UNTIL{\( B_c + b_c \geq B - B_r \)}
   \STATE \( \tilde{\gD} \gets \tilde{\gD} \cup \text{QueryAction}(\pi_v, \tilde{\vmu}, \tilde{\vsigma}, B - B_c) \)
    \STATE \( \pi_{a} \gets \text{BehavioralCloning}(\tilde{\gD}, \pi_{a}, |\tilde{\gD}|) \) with reinitialized \( \pi_a \) 

   \STATE \textbf{return} \( \pi_{a} \)
\end{algorithmic}
\end{small}
\end{algorithm}



\subsection{Stealthy Imitation Countermeasure}
\label{subsec:method_defense}
Although this work focuses on the attacker's perspective, we also propose an effective defense against \acl{SI}. The idea is to leverage the victim's exclusive knowledge of the correct input range; the defender can respond with random actions to invalid queries.
We argue that ignoring queries outside the valid range is not advisable for the victim, as it would leak information about the valid range itself. This approach serves to obfuscate the attacker's efforts to estimate the input range. This defense does not degrade the utility of the victim policy, as it still provides correct answers to valid queries.

%% file: sections/06_Experiment/00_experiments.tex
This section presents our empirical results for \acl{SI}.
We discuss the experimental setup (\Cref{subsec:experimental_setup}), followed by a comparison of our proposed method to baselines (\Cref{subsec:si_vs_baseline}) and analyses and ablation studies (\Cref{subsec:analysis}). Finally, we show the real-world robot policy stealing in \Cref{subsec:panda_robot} and the defense performance in \Cref{subsec:results_defense}.

\subsection{Experimental Setup}
\input{sections/06_Experiment/01_experimental_setup}
\label{subsec:experimental_setup}

\input{sections/06_Experiment/02_results}

%% file: sections/06_Experiment/01_experimental_setup.tex
\noindent\textbf{Victim policies.}
We demonstrate our method on three continuous control tasks from Mujoco~\citep{todorov2012mujoco}: Hopper, Walker2D, and HalfCheetah. The victim policy is trained using \ac{SAC}~\citep{sac}. The victim architecture is a three-layer fully-connected networks (256 hidden units, ReLU activation). The models output a normal distribution from which actions are sampled. These sampled actions are then constrained to the range [-1,1] using tanh. After training, the prediction action given a query state is determined only by the mean of this output distribution. See~\Cref{app:envandvic} for a complete description of all the tasks and performance of the victim policies. \add{All information about compute resources are summarised in \Cref{app:compute}.}

\noindent\textbf{Attacker policies.}
Similar to \citet{Papernot2016TransferabilityIM, knockoff, Orekondy2019PredictionPT}, we employ the architecture of \( \pi_v \) for \( \pi_a \), while omitting the prediction of the standard deviation and incorporating tanh activation.
Our choice of architecture does not significantly influence the refinement of \( S_a \) (see \Cref{app:advarch}), although it does introduce greater variance in the cumulative reward. This phenomenon is attributed to compounding errors, a known issue in imitation learning~\citep{syed2010reduction, ross2011reduction, xu2020error}, where minor training deviations can amplify errors. We set the reserved training budget $B_r = 10^6$ and the base query budget $b_v = 10^5$. Both \( \pi_a \) and \( \pi_e \) share the same architecture and are trained for one epoch per iteration.
We use the Adam optimizer~\citep{Kingma2014AdamAM} with a learning rate of \( \eta = 10^{-3} \) and batch size of 1024.
The final training employs early stopping with a patience of 20 epochs for 2000 total epochs. The reward model \( \hat{\gR} \) is a two-layer fully-connected network (256 hidden neurons, tanh and sigmoid activations). \( \hat{\gR} \) is trained with a learning rate of 0.001 for 100 steps. Prior to training, we apply a heuristic pruning process to $ \mathcal{D}_v $. Specifically, we remove any state-action pairs $ (\vs,\va) $ where any component of $ \va $ equals $ \pm 1 $, corresponding to the maximum and minimal action values. This is due to the tanh activation function in victim policy mapping extreme logit values to these limits, which may not reflect typical decision-making but rather the extremes of the function's output range. This pruning step further assists the reward model in correctly identifying the victim policy's state-action pattern.


\noindent\textbf{Baseline attacks.}
Since our method is the first policy stealing without environment access or prior input range knowledge, we compare it against two approaches: (i) Random: transfer datasets are based on three normal distributions with varying scales, namely \( \mathcal{N}(\mathbf{0}_n,\mathbf{1}_n^2) \), \( \mathcal{N}(\mathbf{0}_n,\mathbf{10}_n^2) \), and \( \mathcal{N}(\mathbf{0}_n,\mathbf{100}_n^2) \); the attacker policy \( \pi_a \) is trained using \ac{BC}; (ii) \ac{DFME}: we adapt the generator-based \ac{DFME}~\citep{dfme} from image classification to control tasks. Convolutional layers are replaced with fully-connected ones, and tanh activation is replaced by batch normalization with affine transformations.

\noindent\textbf{Evaluation.}
We consider two performance metrics: \ac{KL} divergence and return ratio. The \ac{KL} divergence $\KL(S_v \Vert S_a)$ measures the discrepancy between the estimated state distribution $S_a$ and the victim's state distribution $S_v$, an aspect not previously quantified in model stealing. We represent \( S_v \) with a reference normal distribution \( \mathcal{N}(\vmu^*, (\vsigma^*)^2) \) estimated from a dataset of 1 million states, \( \sS_v \), collected from the interaction between the victim policy \( \pi_v \) and the environment. The return ratio, denoted as $rr$, assesses the stealing performance by dividing the average return generated by the attacker policy in the environment by the average return of the victim policy. The return ratio is the average one derived from eight episodes with random initial state. To account for any variability of \ac{SI}, we report results with five random seeds in~\Cref{app:variability_si}.

%% file: sections/06_Experiment/02_results.tex
\subsection{\acl{SI} Attack Performance}
\label{subsec:si_vs_baseline}

\begin{figure*}[ht]
    \centering
    \includegraphics[width=0.8\textwidth]{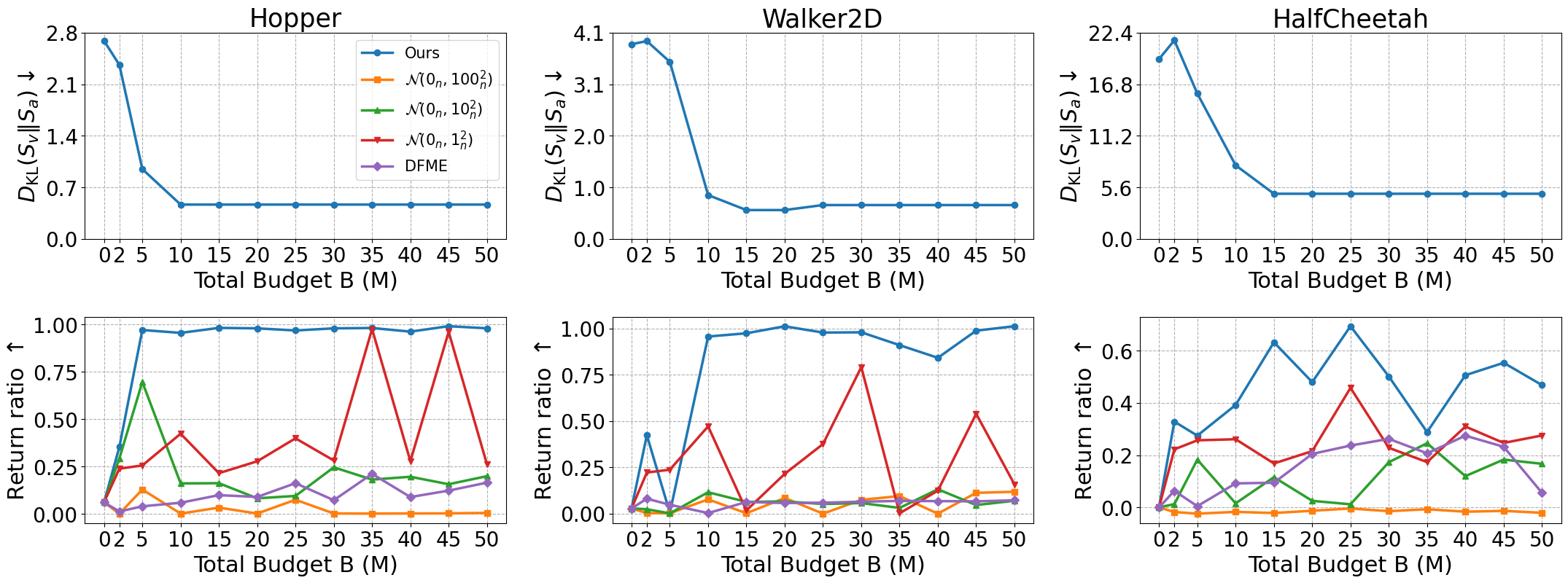}
    \caption{Distribution estimation capacity measured by $\KL(S_v \Vert S_a)$ (top) and return ratio (bottom) as a function of the attacker budget.}
    \label{fig:si_vs_baseline}
\end{figure*}

We assess various policy stealing methods, as shown in~\Cref{fig:si_vs_baseline}.
The measure of $\KL(S_v \Vert S_a)$ is specific to our approach (top row), as the Random strategy does not refine a distribution, and \ac{DFME} focuses on fine-tuning samples. We observe that the gap between \( S_a \) and \( S_v \) becomes consistently smaller and achieves convergence, even when starting from a high value in HalfCheetah. On average, we achieve an 81\% reduction in $\KL(S_v \Vert S_a)$ across all environments. 
Our method substantially outperforms other attacks in terms of return ratio (\Cref{fig:si_vs_baseline}, bottom row). In the Hopper environment, we achieve a return ratio of 97\% with just 5 million queries. In contrast, the best competing method, \(\mathcal{N}(\mathbf{0}_n, \mathbf{10}_n^2)\), under the same query budget reaches only 70\% and quickly falls below 25\%. Further details on the performance of the reward discriminator can be found in~\Cref{app:reward_performance}. While the Random \(\mathcal{N}(\mathbf{0}_n, \mathbf{1}_n^2)\) baseline shows promise in the Hopper environment with 35 million queries, it does not maintain this performance as consistently as ours across varying query budgets. \ac{DFME} fails to steal the victim policy, as it focuses on the near-infinite adversarial samples, restricting exploration. Our method, emphasizing regions instead of individual samples, leads to a more extensive and efficient exploration of the unknown input range. More adaptations of \ac{DFME} are in \Cref{app:adaptation_of_dfme}.



\subsection{Analysis}
\label{subsec:analysis}

\add{\noindent\textbf{Diagonal Gaussian approximation for complex input distributions.} 
The success of our approach, which uses a diagonal Gaussian distribution to approximate real inputs, is robust to complex input distributions.
To support this, we analyze the correlation matrices and the distributions of all variables using real state data from MuJoCo environments. The illustrations in \cref{app:underlying_dist} disclose significant correlations and non-Gaussian distributions among these variables. These findings substantiate the effectiveness of our method in the presence of coupled input variables and complex distributions. Furthermore, in \cref{subsec:panda_robot}, we show a high return ratio when applying our method to more realistically modeled robots with higher-dimensional inputs. This provides additional evidence supporting the applicability of our approach across a broad spectrum of input complexities. In~\Cref{app:full_gaussian_distribution}, we examine the impact of the probabilistic state distribution model and find that a diagonal Gaussian distribution yields better results than a full Gaussian distribution due to the fewer optimized parameters. We also show in~\cref{app:risk_of_exposing_dist} that \ac{RL}-trained policies in control systems can be easily compromised using supervised learning when the input distribution is exposed, even through a diagonal Gaussian distribution. Moreover, \Cref{app:center_scale} experimentally shows that the diagonal Gaussian approximation is robust to estimation errors on $\vmu$ and $\vsigma$.}

\noindent\textbf{Correlation between difficulty of imitation and distribution divergence.}
To empirically evaluate the hypothesis that the difficulty of imitation  is correlated with the divergence between \( S_a \) and \( S_v \), we create 600 estimated state distributions \( S_a \). These distributions are parameterized as \( S_a(\vs; \vz \vsigma^* + \vmu^*, \vsigma^*) \), where each element of \( \vz \) is randomly sampled from a uniform distribution over \([0, 4]\), and its sign is chosen randomly. As a result, the \ac{KL} divergences \( \KL(S_v \Vert S_a) \) for these estimated state distributions range approximately from 0 to 8. For each \( S_a \), we construct a transfer dataset of $10^5$ points and train the attacker's policy \( \pi_a \) using \ac{BC} for one epoch. We measure the average validation loss \( \bar{\Ls}_b \) as a proxy for the difficulty of imitation. We apply Spearman's rank correlation test to these measurements, and the results are summarized in Table~\ref{tab:hypothesis}. These results demonstrate a statistically significant correlation for $\left(\bar{\Ls}_b, \KL\right)$, thus supporting the use of $\pi_e$ as a reliable distribution evaluator in~\Cref{subsec:distribution_estimator}.

\begin{table}
\caption{Spearman's rank correlation between validation loss \( \bar{\Ls}_b \) and distribution divergence \( \KL \).}
\label{tab:hypothesis}
\vskip 0.1in
\begin{center}
\begin{small}

\begin{tabular}{c   S[table-format=-1.2] S[table-format=1.2e-3]}
\toprule
& \multicolumn{2}{c}{$\left(\bar{\Ls}_b, \KL\right)$} \\
\cmidrule(lr){2-3}
\textbf{Environment} & \textbf{Correlation}~\boldmath$\rho$ & \textbf{p-value} \\
\midrule
Hopper & -0.84 & 4.79e-164 \\
Walker2D & -0.78 & 7.59e-122 \\
HalfCheetah & -0.81 & 4.01e-140 \\
\bottomrule
\end{tabular}
\end{small}
\end{center}
\vskip -0.1in
\end{table}

\begin{figure*}
    \centering
    \includegraphics[width=0.75\textwidth]{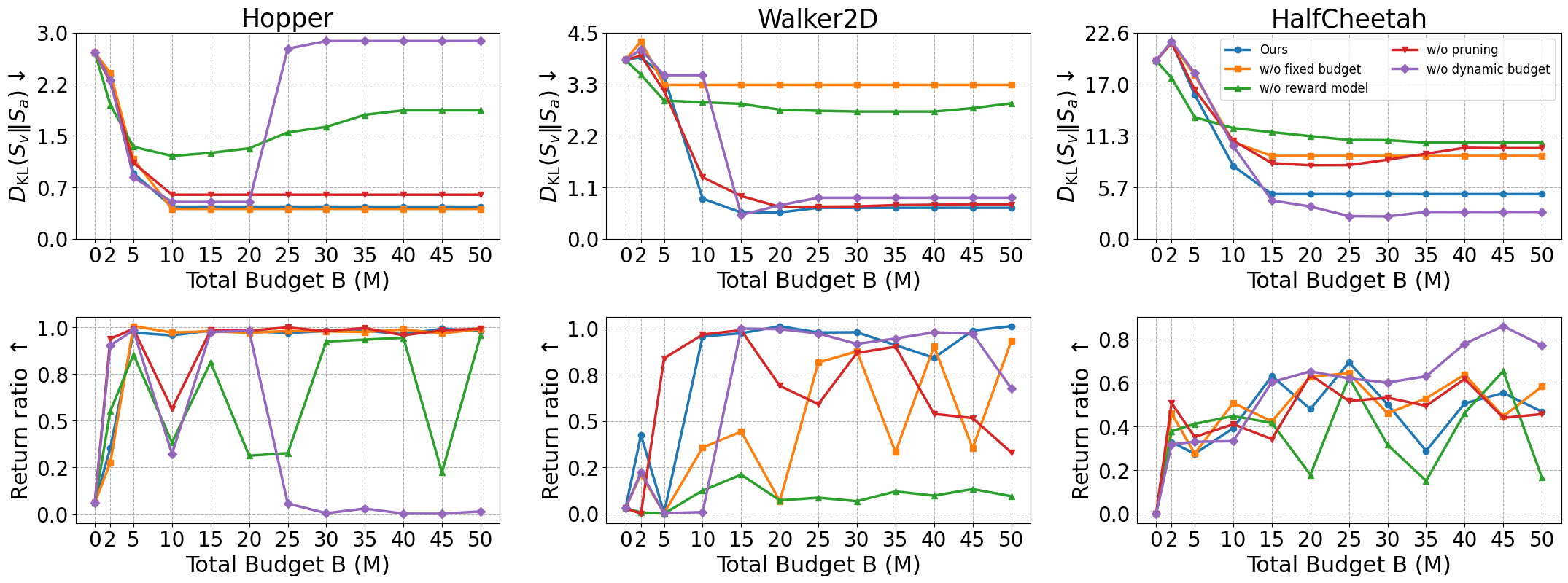}
    \caption{We validate the necessity of (i) fixing the dataset size to train the evaluator model, (ii) dynamic budget, (iii) reward model, and (iv) pruning the transfer dataset.}
    \label{fig:si_vs_ablation}
\end{figure*}

\noindent\textbf{Ablative analysis.}
We study the impact of each component of our method by systematically removing them one at a time, while keeping the other components unchanged. The ablation study includes: (i) the use of $b_v \times \bar{\vsigma}$ instead of $b_v$ samples of transfer dataset $\gD_v$ to train the distribution evaluator $\pi_e$; (ii) bypassing the \textbf{reward model training} and directly using the validation loss $\Ls_b$ of each sample as weight for the \textbf{reward-guided distribution refinement}; (iii) skipping the pruning step of the transfer dataset before training the reward model; and (iv) using $b_v$ instead of $b_v \times \bar{\vsigma}$ to train attacker's policy $\pi_a$ during \textbf{behavioral cloning}. The result is depicted in~\Cref{fig:si_vs_ablation}. We observe that incorporating a reward model can more efficiently minimize the distribution divergence $\KL(S_v \Vert S_a)$. Additionally, employing a fixed budget for the evaluator model helps the attacker select a better \( S_a \), thereby improving the return ratio. We also note the stabilizing effect of pruning the transfer dataset prior to training the reward model. Moreover, if a dynamic budget is not used when constructing the transfer dataset, we observe undesired shifts in \( S_a \) over iterations in Hopper and this leads to a significant reduction in the return ratio.

\subsection{\add{Real-world Robot Policy Stealing}}
\label{subsec:panda_robot}

\begin{figure*}[h]
    \centering
    \includegraphics[width=0.75\textwidth]{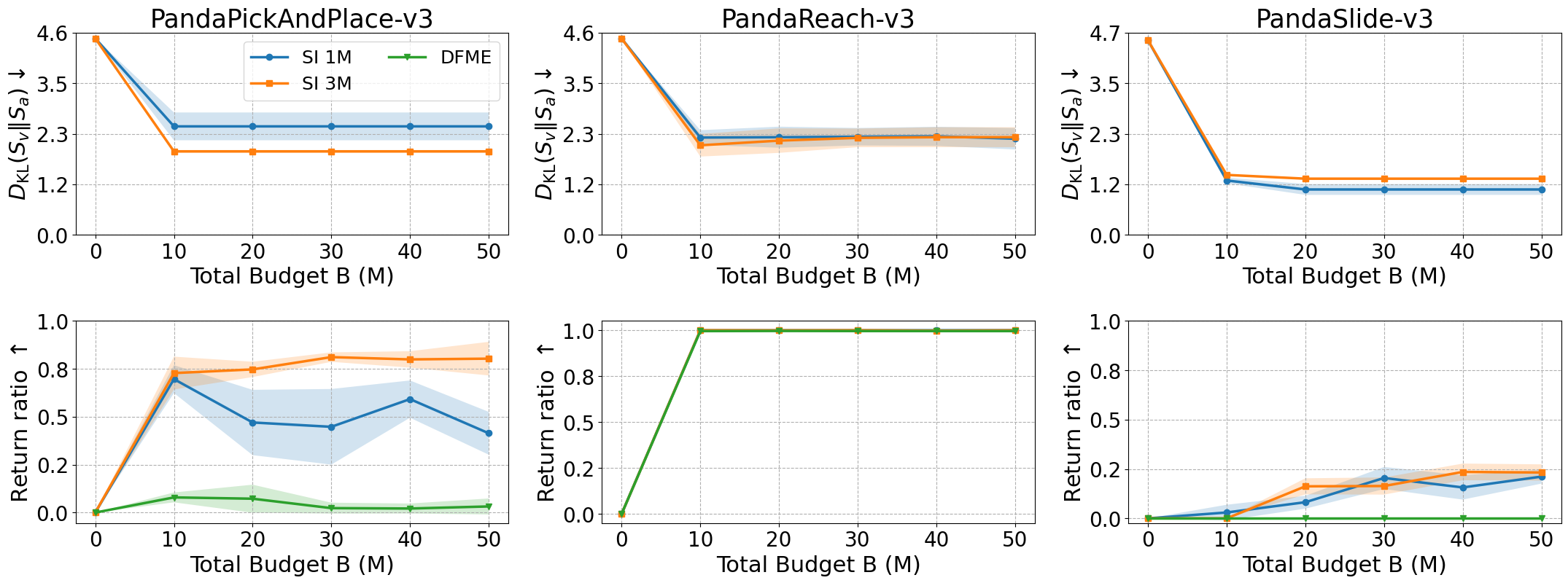}
    \caption{Panda: $\KL(S_v \Vert S_a)$ (top) and return ratio (bottom) as a function of the attacker budget.}
    \label{fig:panda_robot}
\end{figure*}

We validate \acl{SI} in a realistic scenario where the victim policies are trained for the Franka Emika Panda robot simulated by panda-gym~\citep{gallouedec2021pandagym}. The victim policies are from HuggingFace\footnote{https://huggingface.co} and were developed by independent contributors using \ac{TQC}~\citep{tqc}. \ac{TQC} represents a \ac{RL} algorithm distinct from that utilized in Mujoco. The range of returns observed spanned from -50 to 0.


\noindent\textbf{Experimental setup.}
Initializing with $\mathcal{N}(\mathbf{0}_n,\mathbf{1}_n^2)$ by chance leaves little room for optimization, as it already results in a very small $\KL$. To demonstrate efficacy, we initialize the estimated distribution with $\mathcal{N}(\vmu^*+3\vsigma^*,(\vsigma^*)^2)$, thus with all initial $\KL$ being 4.5. The training involves five epochs for the attacker policy per iteration in the loop, with other hyperparameters mirroring those in the Mujoco setup. We calculate the return ratio using $\frac{R(\tau_a)-R_{min}}{R(\tau_v)-R_{min}}$, where $R_{min}$ is the minimal return -50. We conduct the experiments five times, each with a different random seed. To investigate the influence of query budget to reserve after distribution estimation, namely $B_r$, we include an additional experiment in which 3 million queries are reserved.


\noindent\textbf{\acl{SI} results.}
\Cref{fig:panda_robot} illustrates the results, and the task details along with the victim return are summarized in \Cref{app:envandvic}. We observed that \ac{SI} outperforms \ac{DFME} significantly in two of the Panda robot tasks. When the $B_r$ is 1M, labeled as "SI 1M", \ac{SI} achieves a high return ratio with only 10M total budget for PandaPickAndPlace and PandaReach, specifically about $70\%$ and $100\%$ respectively. In the more challenging task of PandaSlide, although the return ratio is only approximately $21\%$, the $\KL$ value significantly decreases by $77\%$. This lower return ratio could be attributed to the unique characteristic of this task: the robot acts mainly at the start, then waits for the object to hit the target. Despite a good distribution estimation, the attacker policy predominantly clones non-essential actions. We also observe improved performance with increased reserved query budgets for PandaPickAndPlace.

\subsection{\acl{SI} Attack Countermeasure}
\label{subsec:results_defense}

We test the efficiency of the proposed defense to our \acl{SI} attack in Mujoco. We consider the input range to match the minimum and maximum values encountered during training.
Upon detecting a query that is outside the predefined input range, the victim policy will uniformly sample an action as a response.
We present the results in~\cref{tab:defense}, evaluating the relative change in \ac{KL} divergence, denoted as $\Delta \KL$, and the return ratio after exhausting the entire 50M query budget. \add{$\Delta \KL$ is calculated as the proportional change from the initial to the final \ac{KL} value, with negative values indicating that the estimated distribution is converging towards the actual distribution.}
The results indicate that the countermeasure substantially impedes the attacker's ability to approximate the victim's distribution, consequently reducing the return ratio of the attacker policy. More elaborations are in \Cref{app:discussion_on_defense}.


\begin{table}
\caption{\add{Results of defense in $\Delta \KL$ and return ratio.}}
\label{tab:defense}
\vskip 0.1in
\begin{center}
\begin{small}

\begin{tabular}{c   c   S[table-format=2.0]   S[table-format=3.0]}
\toprule
\textbf{Environment} &  \textbf{Setup} & \textbf{$\Delta \KL$} & \textbf{Return Ratio} \\
\midrule
\multirow{2}{*}{Hopper} & w/o defense & -83\% & 98\% \\
                        & w/ defense   & +6\% & 0\% \\
\midrule
\multirow{2}{*}{Walker2D} & w/o defense & -83\% & 101\% \\
                          & w/ defense   & +7\% & 7\% \\
\midrule
\multirow{2}{*}{HalfCheetah} & w/o defense & -75\% & 47\% \\
                             & w/ defense   & +9\% & 4\% \\
\bottomrule
\end{tabular}

\end{small}
\end{center}
\vskip -0.1in
\end{table}



%% file: sections/07_Discussion/discussion.tex
\noindent\textbf{Computational efficiency.}
In addition to theft effectiveness, \acl{SI} also demonstrates computational efficiency. The main computational load comes from training the attacker policy $\pi_a$ on the optimized transfer dataset $\hat{\gD}$. This is more computational efficient compared to utilizing all data with size of total budget $B$ like random strategy.

\noindent\textbf{Limitations and future work.}
The limitations in this work present opportunities for future research and exploration. Firstly, attackers should consider the potential effects of initial distribution discrepancies. While our method, initializing the estimated distribution with a standard Gaussian, has proven effective, the threshold beyond which initial distribution divergence compromises effectiveness remains to be identified. Secondly, \acl{SI}, being agnostic to the victim \ac{RL} algorithm, can adapt to various victim policies trained with other \ac{RL} strategies; however, performance may vary across different \ac{RL} algorithms and requires further examination. Finally, expanding our approach to other domains, where acquiring the input range is challenging such as those involving feature vectors, holds considerable promise for future research.

%% file: sections/08_Conclusions/conclusions.tex
We show for the first time that an attacker can successfully steal policies in control systems without requiring environment access or prior knowledge of the input range---a strong attack vector that has not been demonstrated or considered in prior research.
Lacking access to the victim data distribution, we show that a Gaussian assumption for the attacker query data is sufficient for efficient policy extraction.
Our \acl{SI} attack outperforms existing methods adapted to policy stealing for a limited-knowledge attacker.
We show that it is harder to imitate the victim policy when the distribution of the attack queries increasingly aligns that of the victim,
thus allowing an attacker to refine their query distribution.
We encourage policy owners to consider the risks of stealing and to use available defenses, such as the one proposed in this paper, to protect their assets.

%% file: sections/Appendix/01_algorithm.tex


We now provide a detailed description of each function in \Cref{alg:SI}, along with their pseudo code.

\paragraph{Query action.}
Query action (\Cref{alg:query_action}) is the function where we obtain the transfer dataset from victim policy and attacker policy. We sample $b$ state vectors from a Gaussian distribution parameterized by $\vmu$ and $\vsigma$, and obtain responses $\va$ from the policy $\pi$. When $\pi=\pi_v$, the output is dataset $\gD=\gD_v$; otherwise, it is $\gD_a$ when policy is $\pi_a$.

\begin{algorithm}[h!]
\caption{QueryAction}
\label{alg:query_action}
\begin{algorithmic}[1]
   \STATE {\bfseries Input:} Policy $\pi$, mean $\vmu$ and standard deviation $\vsigma$, query budget $b$
   \STATE {\bfseries Output:} Dataset $D$
   \STATE Sample $b$ data points $\vs$ from $\gN(\vmu,\vsigma^2)$
   \STATE $\va \gets \pi(\vs)$
   \STATE $\gD:=\{(\vs_i, \va_i) | i = 1, ..., b\}$
   \STATE \textbf{return} $\gD$
\end{algorithmic}
\end{algorithm}

\paragraph{Behavioral cloning.}
We train policy $\pi$ to mimic the state-action pair mapping in dataset $\gD$ via supervised learning by minimizing the Huber loss, i.e., behavioral cloning in~\Cref{alg:bc}. Considering that the attacker policy $\pi_a$ has different dataset size requirement as distribution evaluator $\pi_e$ using behavioral cloning, we use an additional demand size $N$ to control it.

\begin{algorithm}[h!]
\caption{BehavioralCloning}
\label{alg:bc}
\begin{algorithmic}[1]
   \STATE {\bfseries Input:} Dataset $\gD = \{(\vs_i, \va_i^*)\}$, policy $\pi$, demand size $N$, epochs $E$, learning rate $\eta$
   \STATE {\bfseries Output:} Updated policy $\pi$
   \STATE Sample \( N \) data from \( \mathcal{D} \) and split into training and validation $\gD_t$ and $\gD_v$ 
   \FOR{$e = 1$ to $E$}
      \FOR{each batch $(\vs, \va^*)$ in $\gD_t$}
         \STATE // Compute loss using Huber loss
         \STATE Calculate loss $\Ls_b \gets \text{HuberLoss}(\pi(\vs), \va^*)$
         \STATE // Update model parameters using gradient descent
         \STATE $\pi \gets \pi - \eta \nabla_\pi \Ls_b$
      \ENDFOR
   \ENDFOR
   \STATE \textbf{return} $\pi$
\end{algorithmic}
\end{algorithm}

\paragraph{Train reward.}
We use the code pipeline provided in~\citet{ding} to train the reward model in~\Cref{alg:train_reward}, except for the additional function PruneData. Reward model is trained for total 400 steps in each iteration with learning rate $\eta=10^{-3}$.

\begin{algorithm}[h!]
\caption{TrainReward}
\label{alg:train_reward}
\begin{algorithmic}[1]
   \STATE {\bfseries Input:} dataset $\gD_a$ queried from attacker policy, dataset $\gD_v$ queried from victim policy, reward model $\hat{\gR}$, demand size $N$, total steps $T$, learning rate $\eta$
   \STATE {\bfseries Output:} Trained reward model $\hat{\gR}$
   \STATE Sample \( N \) data from \( \gD_v \) and split into training and validation $\gD_{vt}$ and $\gD_{vv}$ 
   \STATE $\gD'_{vt} \gets \text{PruneData}(\gD_{vt})$
   \FOR{$i = 1$ to $T$}
      \STATE Sample batch data $(\vs_v, \va_v)$ from $\gD'_{vt}$ and $(\vs_a, \va_a)$ from $\gD_a$
      \STATE $L_v \gets -\log(1-\hat{\gR}(\vs_v,\va_v))$
      \STATE $L_a \gets -\log(\hat{\gR}(\vs_a,\va_a))$
      \STATE // Compute the gradient of the total loss
      \STATE $\nabla L \gets \nabla (L_v+L_a)$
      \STATE // Update the reward model
      \STATE $\hat{\gR} \gets \hat{\gR} - \eta \nabla L$
   \ENDFOR
   \STATE \textbf{return} The trained reward model $\hat{\gR}$
\end{algorithmic}
\end{algorithm}

\paragraph{Prune data.}
When the action is equal to maximum or minimal value, i.e., extreme action, it is less likely to be the normal action predicted by the victim policy on the real state distribution, as most control systems do not prefer such extreme action. Extreme action value can easily cause instability in control systems. By pruning the transfer dataset shown in~\Cref{alg:prune_data}, the reward model can identity the difference of state-action pairs coming from the victim and attacker policies. For instance, if there is a state-action pair whose action is an extreme value, then the reward model tends to identity it as a state-action pair from the attacker, as there is no such data in the transfer dataset querying the victim policy after pruning.

\begin{algorithm}[h!]
\caption{PruneData}
\label{alg:prune_data}
\begin{algorithmic}[1]
   \STATE {\bfseries Input:} Dataset $\gD$
   \STATE {\bfseries Output:} Cleaned Dataset $\gD'$
   \STATE $\gD' \gets \emptyset$
   \FOR{each $(\vs_i, \va_i)$ in $\gD$}
      \IF{no element of $\va_i$ equals $1$ or $-1$}
         \STATE $\gD' \gets \gD' \cup \{(\vs_i, \va_i)\}$
      \ENDIF
   \ENDFOR
   \STATE \textbf{return} $\gD'$
\end{algorithmic}
\end{algorithm}
\paragraph{Distribution evaluate.}
The function described in \Cref{alg:distribution_evaluate} is exactly the same as behavioral cloning, but the final output of the function is the validation loss $\bar{\Ls}_b$ of evaluator $\pi_e$.

\begin{algorithm}[h!]
\caption{DistributionEvaluate}
\label{alg:distribution_evaluate}
\begin{algorithmic}[1]
   \STATE {\bfseries Input:} Dataset $\gD = \{(\vs_i, \va_i^*)\}$, policy $\pi$, portion size $N$, epochs $E$, learning rate $\eta$
   \STATE {\bfseries Output:} validation loss $\bar{\Ls}_b$
   \STATE Sample \( N \) data from \( \mathcal{D} \) and split into training and validation $\gD_t$ and $\gD_v$ 
   \FOR{$e = 1$ to $E$}
      \FOR{each batch $(\vs, \va^*)$ in $\gD_t$}
         \STATE Calculate loss $\Ls_b \gets \text{HuberLoss}(\pi(\vs), \va^*)$
         \STATE $\pi \gets \pi - \eta \nabla_\pi \Ls_b$
      \ENDFOR
   \ENDFOR
   \STATE Calculate average validation loss $\bar{\Ls}_b$ on $\gD_v$
   \STATE \textbf{return} $\bar{\Ls}_b$
\end{algorithmic}
\end{algorithm}

\paragraph{Distribution refinement.}
We apply \Cref{eq:distrefine} on the validation split of the transfer dataset to calculate the new $\vmu$ and $\vsigma$, described in~\Cref{alg:dist_refine}.

\begin{algorithm}[h!]
\caption{DistRefine}
\label{alg:dist_refine}
\begin{algorithmic}[1]
   \STATE {\bfseries Input:} dataset $\gD$, reward model $\hat{\gR}$, demand size $N$
   \STATE {\bfseries Output:} updated $\vmu'$ and $\vsigma'$
   \STATE Sample \( N \) data from \( \gD \) and split into training and validation $\gD_{t}$ and $\gD_{v}$ 
   \STATE $\gD'_{v} \gets \text{PruneData}(\gD_{v})$
   \STATE $\vmu' \gets \frac{\sum_{(\vs, \va) \in \gD'_{v}} \hat{r}(\vs,\va) \cdot \vs}{\sum_{(\vs, \va) \in \gD'_{v}} \hat{r}(\vs,\va)}$
   \STATE $\vsigma'^2 \gets \frac{\sum_{(\vs, \va) \in \gD'_{v}} \hat{r}(\vs,\va) \cdot (\vs-\vmu')^2}{\sum_{(\vs, \va) \in \gD'_{v}} \hat{r}(\vs,\va)}.$
   \STATE $\vsigma'=\sqrt{\vsigma'^2}$
   \STATE \textbf{return} $\vmu'$ and $\vsigma'$
\end{algorithmic}
\end{algorithm}

%% file: sections/Appendix/02_EnvAndVictim.tex
We conducted our experiments on environments sourced from Gymnasium~\citep{towers_gymnasium_2023}. The specific environments, along with their version numbers and the performance metrics of the victim policies, are detailed in~\Cref{tab:env_vic}. The victim policies are trained using the Ding repository~\citep{ding}, a reputable source for PyTorch-based RL implementations~\citep{paszke2017automatic}. We employ \ac{SAC} to train the victim policy; hence, the victim policy comprises an actor and a critic model. The actor model receives the state as input and outputs the action distribution, while the critic model receives a concatenated state and action as input and outputs the Q-value. During queries to the victim policy, only the actor model is utilized, outputting the mean of the action distribution as a response. The state observations primarily consist of the positional coordinates and velocities of various body parts. The video of victim policy is in supplementary material.

\begin{table}[h!]
\caption{Mujoco environments and performance of victim policy.}
\label{tab:env_vic}
\vskip 0.1in
\begin{center}
\begin{tabular}{cccc}

\toprule
\textbf{Environment}  & \textbf{Observation space} & \textbf{Action space} & \textbf{Victim return} 
\\
\midrule
Hopper-v3      & 11                         & 3                     & 3593$\pm$3            \\
Walker2D-v3    & 17                         & 6                     & 4680$\pm$43           \\
HalfCheetah-v3 & 17                         & 6                     & 12035$\pm$61 \\
\bottomrule
\end{tabular}
\end{center}
\vskip 0.1in
\end{table}

For the victim policies in the Panda robot stealing setup, we obtain them directly from HuggingFace, instead of training them ourselves, to simulate a real-world threat. The performance of these victim policies is described in \Cref{tab:panda_robot}.

\begin{table}[h!]
\caption{Panda environments and performance of victim policy.}
\label{tab:panda_robot}
\vskip 0.1in
\begin{center}

\begin{tabular}{cccc}
\toprule
\textbf{Environment} & \textbf{Observation space} & \textbf{Action space} & \textbf{Victim return} \\
\midrule
PandaPickAndPlace-v3 & 25 & 4 & -7$\pm$4 \\
PandaReach-v3 & 12 & 3 & -2$\pm$1 \\
PandaSlide-v3 & 24 & 3 & -12$\pm$7 \\
\bottomrule
\end{tabular}

\end{center}
\vskip -0.1in
\end{table}

%% file: sections/Appendix/15_compute.tex
In this section, we provide detailed information on the computational resources used for our main experiments. All experiments were conducted on a single NVIDIA GeForce RTX 2080 Ti GPU. The time required to train the victim policies within the Mujoco environment varied depending on the scenario. Specifically, the Hopper and HalfCheetah models were trained in approximately 12 hours, while the Walker2D model required a more extensive duration, taking up to 2 days. \acl{SI} completed within 2 hours, irrespective of the query budget. In contrast, the Random strategy's compute time varied from 1 to 12 hours, based on the query budget. This variability arises from the Random strategy's approach of using the entire dataset to train the attacker policy, unlike our method, which only uses the most effective dataset. For \ac{DFME}, the compute time is linearly related to the query budget, with a completion time of 2 hours at a 50M query budget.
Regarding the Panda task, the compute time is approximately 4 hours for each query budget checkpoint.

%% file: sections/Appendix/03_ChoiceArch.tex
We investigate the impact of various attacker policy architectures on performance when executing \acl{SI}. Each victim policy utilizes a three-layer fully-connected network with 256 hidden units. To understand the effect of architecture variations, we modify the attacker policies by adjusting the layer numbers to 4, 6, and 10. Furthermore, we conduct experiments with the original layer structure, but reduce the hidden units to 128.

\begin{figure}[h!]
    \centering
    \includegraphics[width=0.85\textwidth]{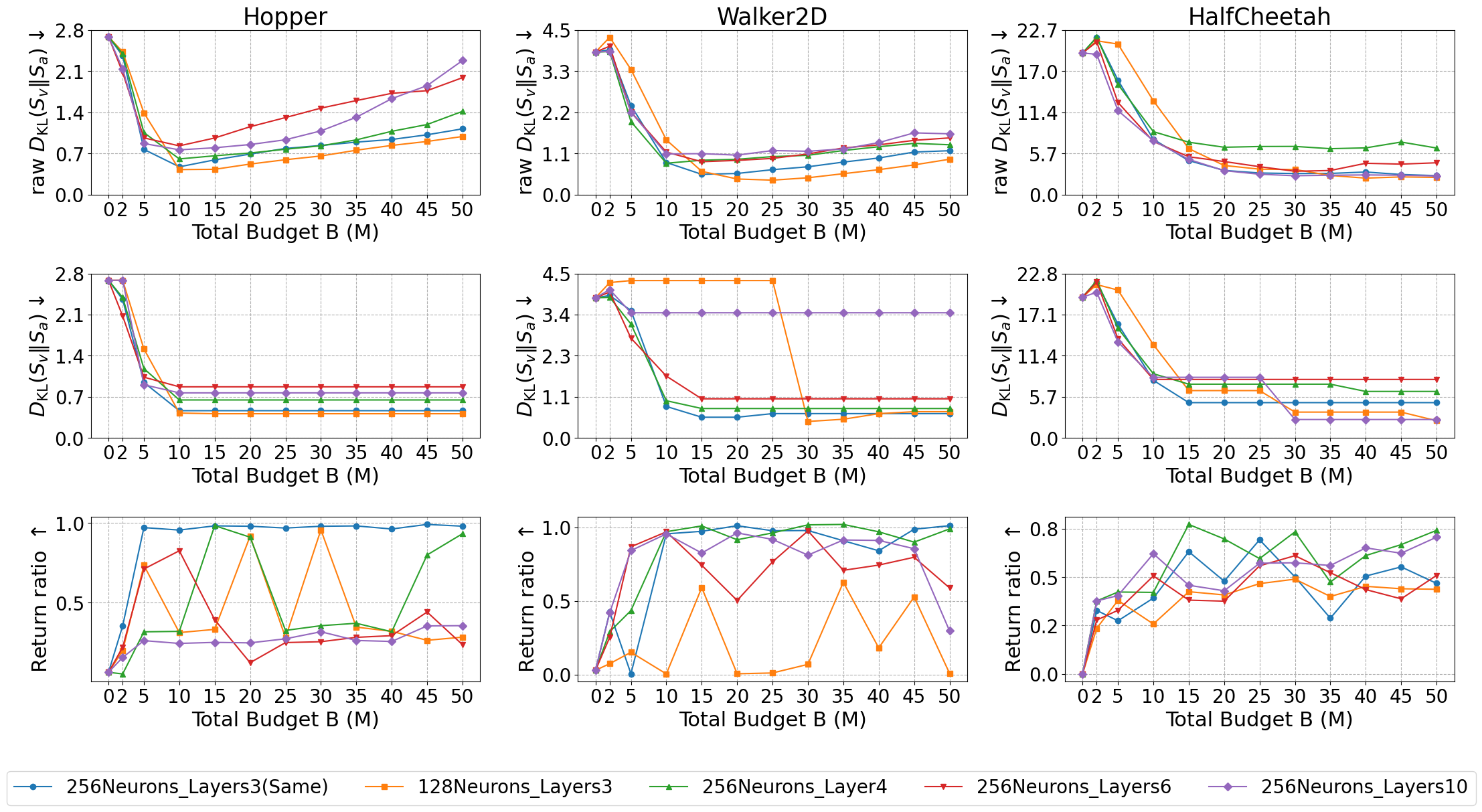}
    \caption{Influence of model architecture on stealing performance.}
    \label{fig:si_vs_arch}
\end{figure}

We depict the results on~\Cref{fig:si_vs_arch}. To better understand the impact, except for $\KL(S_v \Vert S_a)$ and return ratio, we also provide raw $\KL(S_v \Vert S_a)$ on top row, which is the last $\KL(S_v \Vert S_a)$ at the end of the iteration, rather than the one selected by distribution evaluator $\pi_e$. We observe that the raw $\KL$ of different architecture choices exhibit similar tendencies, thus the architecture choice has limited impact on the distribution refinement. In the second row of \Cref{fig:si_vs_arch}, except for Walker2d, the selection of the $\KL(S_v \Vert S_a)$ during refining by $\pi_e$ guarantee an appropriate estimated distribution $S_a$ and low $\KL(S_v \Vert S_a)$, preventing the divergence of distribution approximation. However, we observe that the return ratio exhibits higher variance in the third row. This indicates that the return ratio is sensitive when the architecture is different, even when the estimated distribution is closed to the real state distribution. This is also a challenge in the realm of imitation learning, known as compounding errors~\citep{syed2010reduction, ross2011reduction, xu2020error}. Compounding errors imply that even minor training errors can snowball into larger decision errors. In our case, the minor training error comes from different architecture choices.

It is essential to highlight that this issue of compounding errors is predominantly absent in image classification model stealing, where test data points are independently evaluated. Nonetheless, the robustness of the estimation of the underlying distribution $S_v$ in terms of KL divergence underscores the effectiveness of our approach.

%% file: sections/Appendix/05_Variability_of_Stealthy_Imitation.tex
\begin{figure}[]
    \centering
    \includegraphics[width=0.75\textwidth]{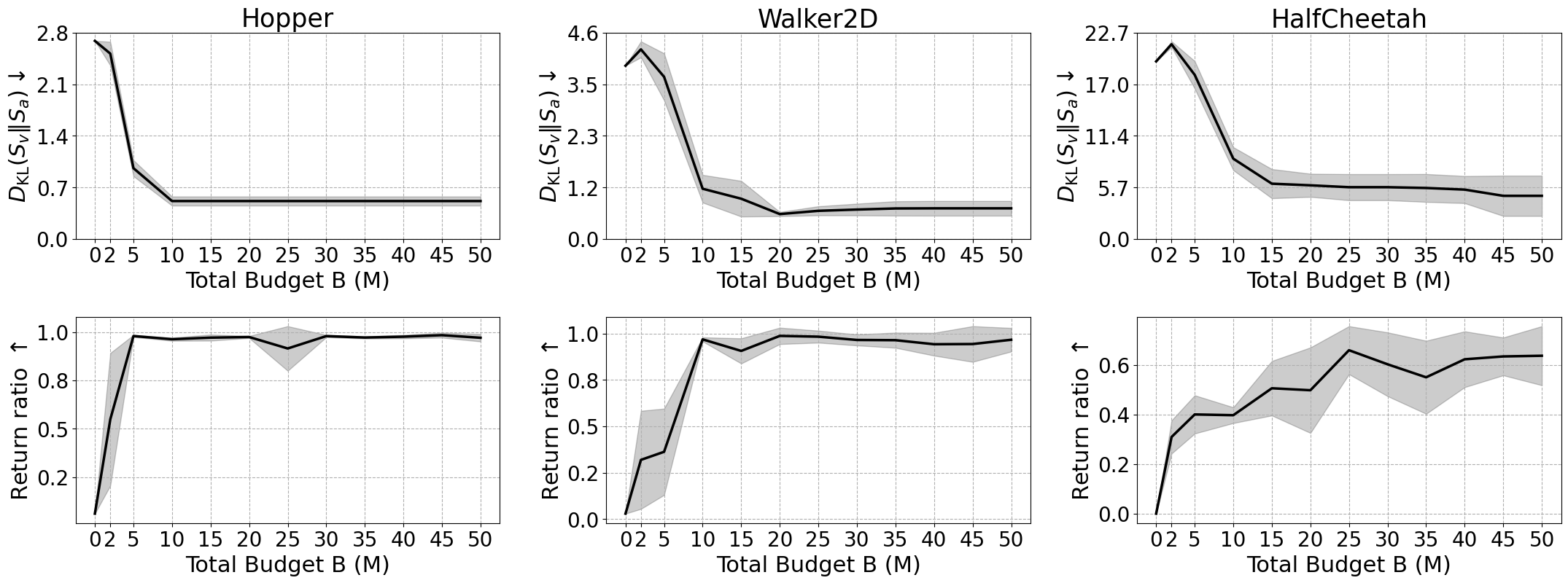}
    \caption{Variability of policy stealing performances.}
    \label{fig:errorbar_plot}
\end{figure}

We report the variability of \acl{SI} in \Cref{fig:errorbar_plot} by using five random seeds to obtain five estimated distributions $S_a$ and train five attacker policies $\pi_a$. The performance of each policy is still obtained by collecting the average return ratio from eight episodes. We observe that the variability of $\KL(S_v \Vert S_a)$ has impact on that of the return ratio, suggesting that a reliable estimated distribution is crucial to attacker policy training. The exact experimental results of Mujoco are listed in~\Cref{tab:kl_rr_mean_std_mujoco} and Panda in~\Cref{tab:kl_rr_mean_std_panda}

\begin{table}[h]
\caption{$\KL(S_v \Vert S_a)$ and return ratio across Mujoco environments with different total budget using Stealthy Imitation.}
\label{tab:kl_rr_mean_std_mujoco}
\vskip 0.1in
\begin{center}
\begin{tabular}{ccccc}
\toprule
\textbf{Metric} & \textbf{Total budget B(M)} & \textbf{Hopper} & \textbf{Walker2D} & \textbf{Halfcheetah} \\
\midrule
\multirow{12}{*}{$\KL(S_v \Vert S_a)$} & 0 & $2.68\pm0.00$ & $3.87\pm0.00$ & $19.55\pm0.00$ \\
 & 2 & $2.51\pm0.16$ & $4.23\pm0.18$ & $21.44\pm0.32$ \\
& 5 & $0.96\pm0.11$ & $3.62\pm0.53$ & $18.07\pm1.52$ \\
 & 10 & $0.51\pm0.06$ & $1.12\pm0.31$ & $8.83\pm1.27$ \\
  & 15 & $0.51\pm0.06$ & $0.90\pm0.40$ & $6.08\pm1.61$ \\
 & 20 & $0.51\pm0.06$ & $0.55\pm0.05$ & $5.90\pm1.27$ \\
  & 25 & $0.51\pm0.06$ & $0.63\pm0.10$ & $5.69\pm1.44$ \\
 & 30 & $0.51\pm0.06$ & $0.65\pm0.13$ & $5.69\pm1.44$ \\
  & 35 & $0.51\pm0.06$ & $0.68\pm0.16$ & $5.60\pm1.54$ \\
 & 40 & $0.51\pm0.06$ & $0.68\pm0.16$ & $5.42\pm1.49$ \\
  & 45 & $0.51\pm0.06$ & $0.68\pm0.16$ & $4.74\pm2.22$ \\
 & 50 & $0.51 \pm0.06$ & $0.68\pm0.16$ & $4.74\pm2.22$ \\
\midrule
\multirow{12}{*}{Return ratio (\%)} & 0 & $6.15\pm0.00$ & $2.87\pm0.00$ & $0.00\pm0.00$ \\
 & 2 & $54.89\pm34.25$ & $31.87\pm26.44$ & $30.96\pm6.70$ \\
& 5 & $97.90\pm0.55$ & $36.21\pm23.27$ & $40.05\pm7.72$ \\
 & 10 & $96.25\pm0.80$ & $96.81\pm1.10$ & $39.75\pm3.18$ \\
  & 15 & $97.07\pm1.57$ & $90.59\pm6.79$ & $50.62\pm11.00$ \\
 & 20 & $97.39\pm0.53$ & $98.69\pm4.42$ & $49.83\pm17.26$ \\
  & 25 & $91.47\pm11.51$ & $98.27\pm3.23$ & $65.98\pm9.69$ \\
 & 30 & $97.87\pm0.56$ & $96.50\pm2.95$ & $60.30\pm12.87$ \\
  & 35 & $97.13\pm0.58$ & $96.37\pm4.09$ & $55.04\pm14.75$ \\
 & 40 & $97.57\pm0.90$ & $94.20\pm6.15$ & $62.32\pm11.27$ \\
  & 45 & $98.33\pm1.28$ & $94.31\pm9.61$ & $63.47\pm7.63$ \\
 & 50 & $97.00 \pm1.92$ & $96.61\pm6.35$ & $63.74\pm11.92$ \\

\bottomrule
\end{tabular}
\end{center}
\vskip -0.1in
\end{table}

\begin{table}[h]
\caption{$\KL(S_v \Vert S_a)$ and return ratio across Panda environments with different reserved budget and total budget using Stealthy Imitation.}
\label{tab:kl_rr_mean_std_panda}
\vskip 0.1in
\begin{center}
\begin{small}
\begin{tabular}{cccccc}
\toprule
\textbf{Metric} & \textbf{Reserved budget(M)} & \textbf{Total budget(M)} & \textbf{PandaPickAndPlace-v3} & \textbf{PandaReach-v3} & \textbf{PandaSlide-v3} \\
\midrule
\multirow{12}{*}{$\KL(S_v \Vert S_a)$} & \multirow{6}{*}{1} & 0 & $4.50\pm0.00$ & $4.50\pm0.00$ & $4.50\pm0.00$ \\
 & & 10 & $2.48\pm0.32$ & $2.23\pm0.17$ & $1.26\pm0.06$ \\
 & & 20 & $2.48\pm0.32$ & $2.24\pm0.24$ & $1.05\pm0.12$ \\
 & & 30 & $2.48\pm0.32$ & $2.25\pm0.20$ & $1.05\pm0.12$ \\
 & & 40 & $2.48\pm0.32$ & $2.26\pm0.22$ & $1.05\pm0.12$ \\
 & & 50 & $2.48\pm0.32$ & $2.20\pm0.25$ & $1.05\pm0.12$ \\
 \cmidrule{2-6}
 & \multirow{6}{*}{3} & 0 & $4.50\pm0.00$ & $4.50\pm0.00$ & $4.50\pm0.00$ \\
 & & 10 & $1.91\pm0.03$ & $2.05\pm0.26$ & $1.39\pm0.00$ \\
 & & 20 & $1.91\pm0.03$ & $2.16\pm0.28$ & $1.30\pm0.00$ \\
 & & 30 & $1.91\pm0.03$ & $2.22\pm0.21$ & $1.30\pm0.00$ \\
 & & 40 & $1.91\pm0.03$ & $2.24\pm0.23$ & $1.30\pm0.00$ \\
 & & 50 & $1.91\pm0.03$ & $2.24\pm0.23$ & $1.30\pm0.00$ \\
\midrule
\multirow{12}{*}{Return ratio (\%)} & \multirow{6}{*}{1} & 0 & $0.00\pm0.00$ & $0.00\pm0.00$ & $0.00\pm0.00$ \\
 & & 10 & $69.61\pm7.27$ & $99.99\pm0.08$ & $2.98\pm3.91$ \\
 & & 20 & $46.99\pm17.03$ & $100.00\pm0.07$ & $8.18\pm3.26$ \\
 & & 30 & $44.73\pm19.72$ & $99.99\pm0.08$ & $20.45\pm5.57$ \\
 & & 40 & $59.18\pm9.72$ & $100.03\pm0.08$ & $15.65\pm6.15$ \\
 & & 50 & $41.38\pm11.16$ & $100.05\pm0.08$ & $21.14\pm3.55$ \\
  \cmidrule{2-6}
 & \multirow{6}{*}{3} & 0 & $0.00\pm0.00$ & $0.00\pm0.00$ & $0.00\pm0.00$ \\
 & & 10 & $72.75\pm8.57$ & $100.04\pm0.04$ & $0.00\pm0.00$ \\
 & & 20 & $74.67\pm3.99$ & $99.99\pm0.07$ & $16.22\pm4.08$ \\
 & & 30 & $81.08\pm2.46$ & $100.04\pm0.12$ & $16.38\pm4.34$ \\
 & & 40 & $79.88\pm4.29$ & $99.84\pm0.07$ & $23.52\pm4.22$ \\
 & & 50 & $80.28\pm8.77$ & $99.99\pm0.08$ & $23.22\pm4.14$ \\
\bottomrule
\end{tabular}
\end{small}
\end{center}
\vskip -0.1in
\end{table}


%% file: sections/Appendix/09_reward_model_loss.tex
\begin{figure}[h!]
    \centering
    \includegraphics[width=1\textwidth]{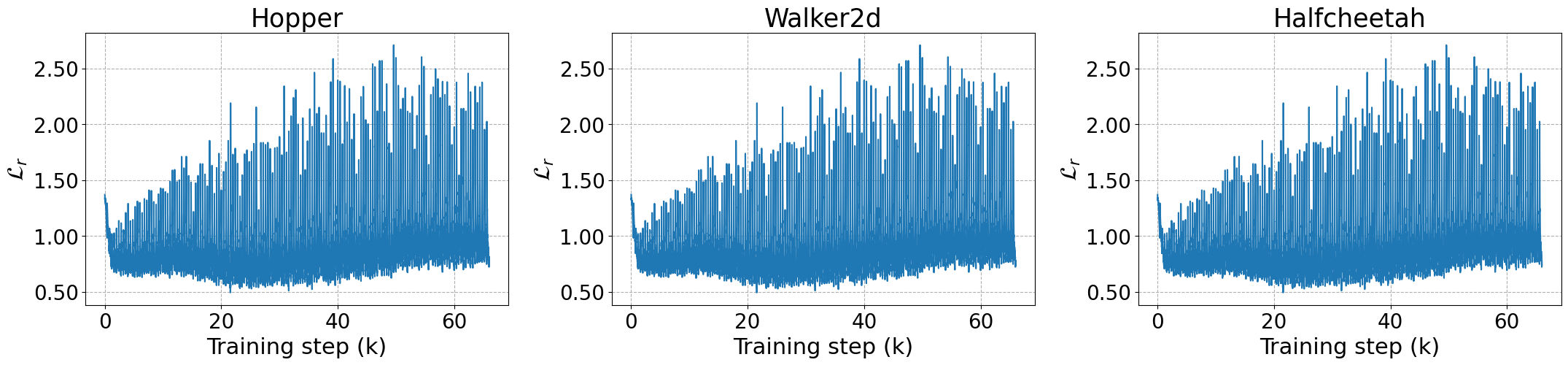}
    \caption{The reward discriminator loss in~\Cref{eq:loss_reward}.}
    \label{fig:reward_model_loss}
\end{figure}

In this section, we analyze how the reward discriminator loss defined in \Cref{eq:loss_reward} changes throughout the distribution estimation process (\Cref{fig:reward_model_loss}). In each iteration, we train the reward model for 400 steps; in each step, a batch of data will be sampled from the current victim and attacker distributions, $\gD_v$ and $\gD_a$ respectively. The x axis in~\Cref{fig:reward_model_loss} represents the number of steps using a total of 50 million query budget. 

We observe that the reward discriminator exhibits oscillations with the variation of the estimated distribution and attacker policy through the iterations. The discriminator's loss may decrease when it successfully identifies attacker's state and action pair data but can increase again as the estimated distribution shifts to a new region where the reward model has not been trained.

%% file: sections/Appendix/14_DFME.tex
\acl{SI} is the first method to steal a policy without the knowledge of input range, which means that there are no baselines for this setup. However, we want to provide a comparison to prior art and adapt \ac{DFME} to the best of our ability to the present setup.

\begin{figure}[h!]
    \centering
    \includegraphics[width=0.8\textwidth]{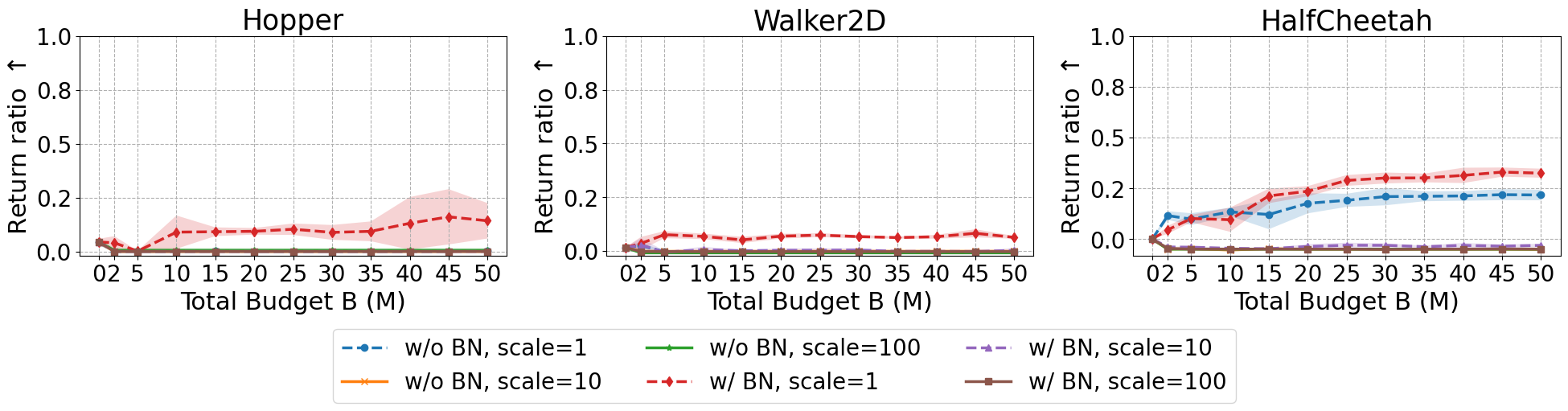}
    \caption{Attacker policy return ratios achieved by adapting \ac{DFME} with modifications: batch normalization (w/ BN) and without (w/o BN), along with output scaling by factors of 1, 10, and 100. Experiments were conducted using five distinct random seeds.}
    \label{fig:adaptation_dfme}
\end{figure}

The original \ac{DFME} generator used a tanh activation function, confining outputs to [-1,1], typical in image classification model stealing. 
We modified the generator in \ac{DFME} by either substituting tanh with batch normalization incorporating learnable scaling and shift factors (w/ BN) to enable exploration beyond the initial \( \mathcal{N}(\mathbf{0}_n, \mathbf{1}_n^2) \), or by removing tanh without batch normalization (w/o BN). Additionally, we expanded the initial state range by scaling the generator's output by factors of 1, 10, and 100. We conduct the experiments with five different random seeds and report results in~\cref{fig:adaptation_dfme}.

We find that all attacker policies yield low return ratios, showing \ac{DFME}'s inadequacy in scenarios with unknown input ranges. This stems from \ac{DFME}'s limitation: in any initial state distribution, such as \( \mathcal{N}(\mathbf{0}_n, \mathbf{1}_n^2) \), it consistently synthesizes samples, where victim and attacker policies disagree. The infinite amount of adversarial samples restrict its ability to explore distributions with varying means and scales. In contrast, our \acl{SI} method uniquely tackles such problem by analyzing entire regions through a reward model, which evaluates overall regional performance rather than individual samples. We outline the differences between \ac{DFME} and \acl{SI} in~\Cref{tab:comparison_dfme_si}, highlighting how each method addresses distinct threat models and objectives.

\begin{table}[h!]
\caption{Contrasting DFME and SI in high level regarding to the query data, target, focus, reusability, and advantage.}
\label{tab:comparison_dfme_si}
\vskip 0.1in
\begin{center}
\begin{tabular}{ccc}

\toprule
\textbf{Type}  & \textbf{DFME} & \textbf{SI} 
\\
\midrule
Query data         & Model-generated                                        & Probability distribution                          \\
Target              & Adversarial examples                                   & Hard-to-Imitate regions                           \\
Focus               & Enhancing sample difficulty using $L_1$ loss          & Assessing overall difficulty via a reward model   \\
Reusability         & Limited, dependent on the model                        & High, as the state distribution is consistent     \\
Advantage           & Improves data distribution with input range insight   & Effectively determines input range                \\
\bottomrule
\end{tabular}
\end{center}
\vskip 0.1in
\end{table}

%% file: sections/Appendix/04_varying_center_scale.tex
\begin{figure}[!htbp]
    \centering
    \includegraphics[width=0.65\textwidth]{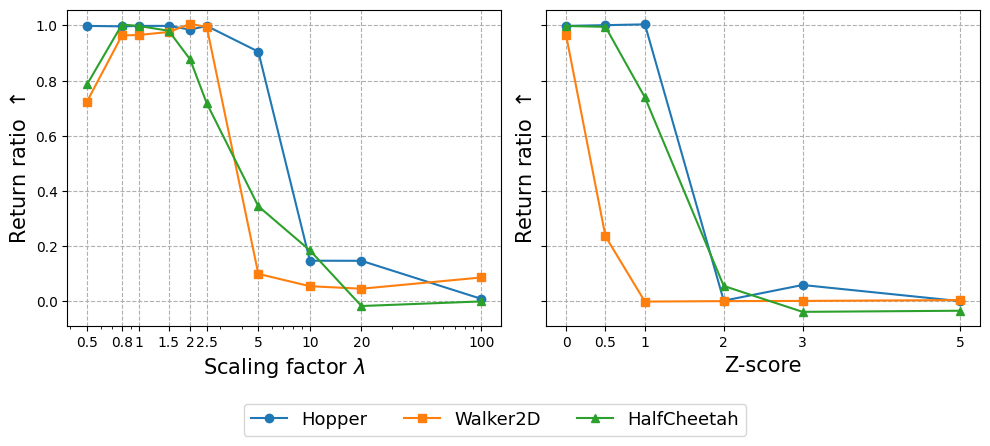}
    \caption{Left: policy stealing performance (return ratio) when \( \vmu = \vmu^* \) and the scale factor \( \lambda \) modifies \( \vsigma^* \) such that \( S_a=\mathcal{N}(\vmu^*, (\lambda \vsigma^*)^2) \). Right: policy stealing performance (return ratio) with \( \vsigma = \vsigma^* \) and \( \vmu = \vz \vsigma^* + \vmu^* \), such that \( S_a=\mathcal{N}(\vz\vsigma^*+\vmu^*, (\vsigma^*)^2) \).}
    \label{fig:combined_performance_zscore_lambda}
\end{figure}

We customize $S_a$ with different parameters to explore the effect of discrepancy between $S_a$ and $S_v$. The left of~\Cref{fig:combined_performance_zscore_lambda} explores the impact of varying \( \vsigma \) while holding \( \vmu = \vmu^* \) constant such that \( S_a=\mathcal{N}(\vmu^*, (\lambda \vsigma^*)^2) \) with a factor $\lambda$. Conversely, the right investigates the effect of modifying \( \vmu \) while keeping \( \vsigma = \vsigma^* \) constant, \( S_a=\mathcal{N}(\vz\vsigma^*+\vmu^*, (\vsigma^*)^2) \). Different values of \( \vz \) serve as a measure of the divergence between the estimated \( \vmu \) and \( \vmu^* \). The sign of each element in $\vz$ is randomly chosen. Transfer datasets, each containing 1 million queries, are generated from these customized distributions. These datasets are then used to train the attacker's policy \( \pi_a \) through \ac{BC} for up to 2000 epochs, utilizing early stopping with a patience of 20 epochs. From \Cref{fig:combined_performance_zscore_lambda} we observe that minor variations in \( \vsigma \) are more tolerable compared to deviations in \( \vmu \).

%% file: sections/Appendix/13_risk_of_exposing_dist.tex
\begin{figure*}
    \centering
    \includegraphics[width=0.81\textwidth]{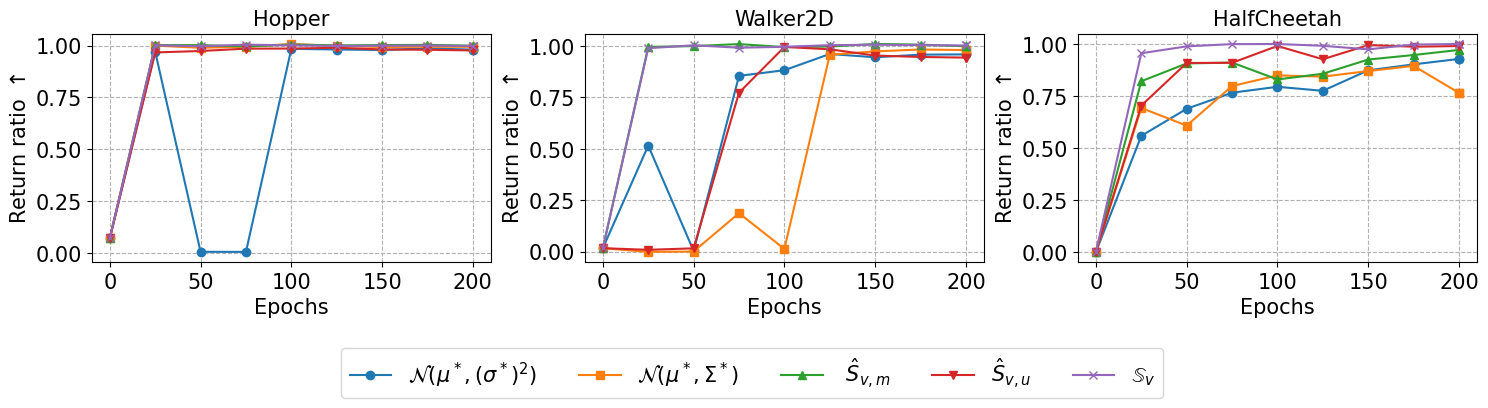}
    \caption{Model stealing success for different choices of \( S_a \) based on the underlying distribution \( S_v \).}
    \label{fig:risk_of_exposing_dist}
\end{figure*}

To demonstrate the risk of exposing the input distribution, we train $\pi_a$ via \acl{BC} for 200 epochs on five different distributions for \( S_a \), each approximated directly from the real state dataset \( \sS_v \): (i) and (ii) \( \mathcal{N}(\vmu^*, (\vsigma^*)^2) \) and \( \mathcal{N}(\vmu^*, \mSigma^*) \): the mean \( \vmu^* \) and variance \( (\vsigma^*)^2 \) or covariance \( \mSigma^* \) are directly calculated from \( \sS_v \), representing diagonal and full covariance matrix, respectively;
(iii) and (iv) \( \hat{S}_{v,u} \) and \( \hat{S}_{v,m} \): these are non-parametric distribution approximations derived using \ac{KDE}, treating variables as independent and dependent, respectively; and
(v) \( \sS_v \): This samples data directly from the real states. \Cref{fig:risk_of_exposing_dist} shows that successful policy stealing is feasible even when queries are sampled from an approximate distribution, even through a diagonal Gaussian distribution.

%% file: sections/Appendix/11_discussion_on_defense.tex
In this section, we further elaborate on the defense method proposed in~\Cref{subsec:method_defense}. We begin by explaining why the defense works and provide potential strategies for both defenders and attackers, especially when attackers become aware of our defense tactics. We then summarize the key lessons for defenders.
\paragraph{Why does the defense work?}
\ac{SI} is used to estimate the real distribution by observing the difficulty attackers face in mimicking actions across various estimated distributions. A simple yet effective defense against \ac{SI} involves randomizing outputs for states outside a known input range, thereby increasing the difficulty in distinguishing between different estimated distributions.
\paragraph{How to enhance the defense?}
If attackers realize this defense strategy and begin to identify the randomness by analyzing output variance, they might develop new methods of attack. In response, our defense can be enhanced to produce similar variance for both in-range and out-of-range queries. For example, mapping out-of-range queries to a random or fixed point within the range could yield consistent output variance.
\paragraph{What does \ac{SI} teach us as defenders?}
Defenders should be wary of revealing their true input distributions. Such disclosure could potentially expose \ac{RL} trained policies to the risk of policy stealing via supervised learning.

%% file: sections/Appendix/18_full_gaussian_distribution.tex
\begin{figure}[h]
    \centering
    \includegraphics[width=0.75\textwidth]{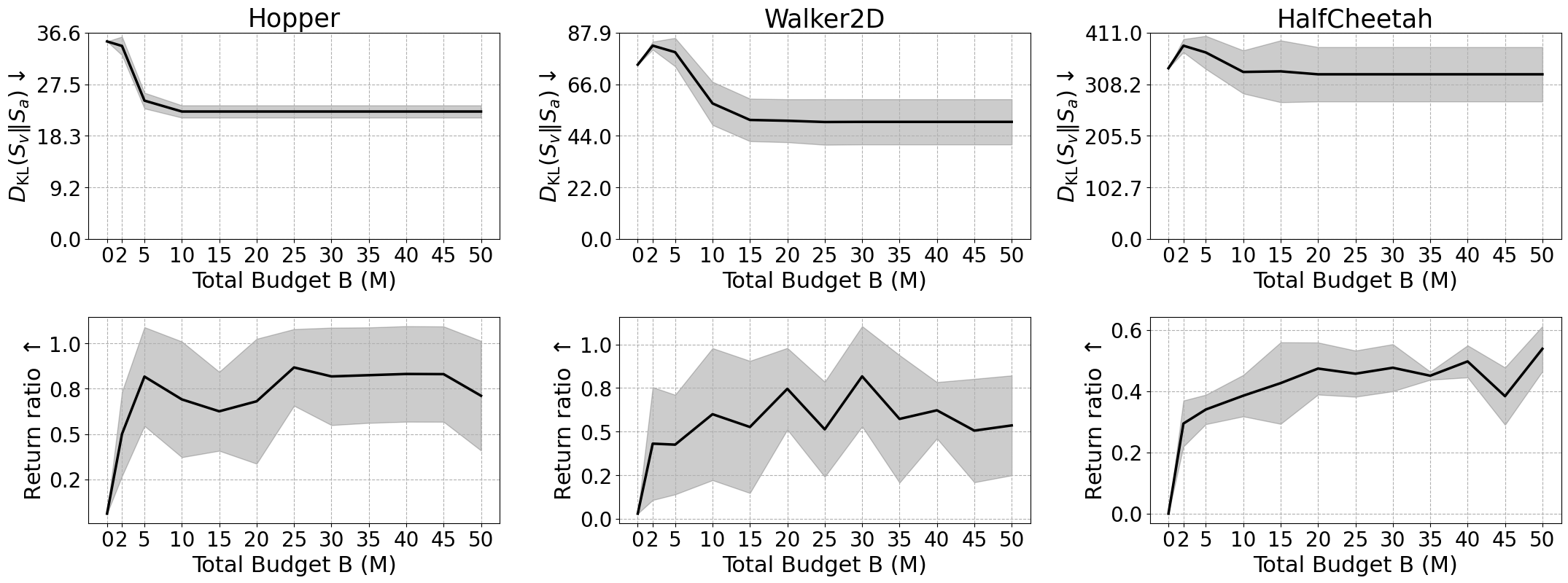}
    \caption{The results of the distribution divergence and the performance of the attacker policy when the estimated distribution is initialized with a full Gaussian distribution and optimized in Stealthy Imitation.}
    \label{fig:full_gaussian_distribution}
\end{figure}

To investigate the impact of the probabilistic state distribution model, we initialize the estimated distribution with not only a diagonal Gaussian distribution but also a full Gaussian distribution, and optimize it in Stealthy Imitation. The experiments are repeated five times with different random seeds. As shown in~\Cref{fig:full_gaussian_distribution}, our method proves effective with different estimated distributions, though it may result in reduced return ratios and increased variance compared with a diagonal Gaussian distribution in~\Cref{fig:errorbar_plot}. This difference primarily stems from the fewer parameters of the diagonal Gaussian, which simplifies the optimization process.

%% file: sections/Appendix/12_underlying_distribution.tex
\begin{figure}[ht]
    \centering
    \includegraphics[width=1\textwidth]{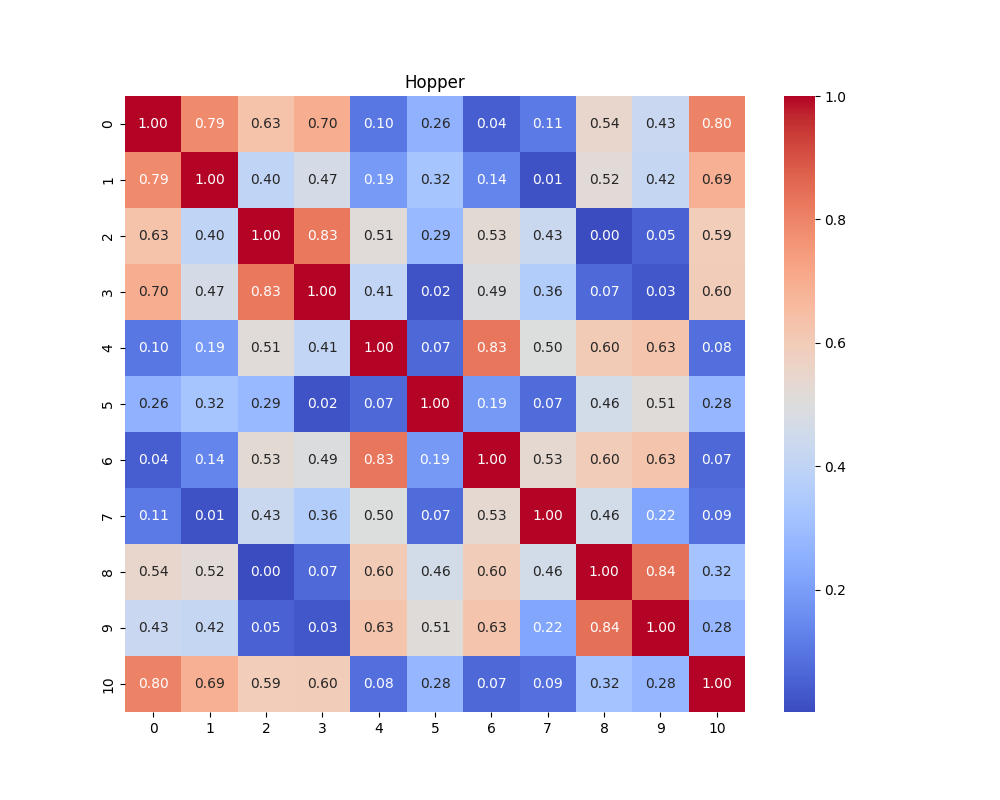}
    \caption{Correlation matrix of the states collected from Hopper environment.}
    \label{fig:cm_hopper}
\end{figure}
\begin{figure}[ht]
    \centering
    \includegraphics[width=1\textwidth]{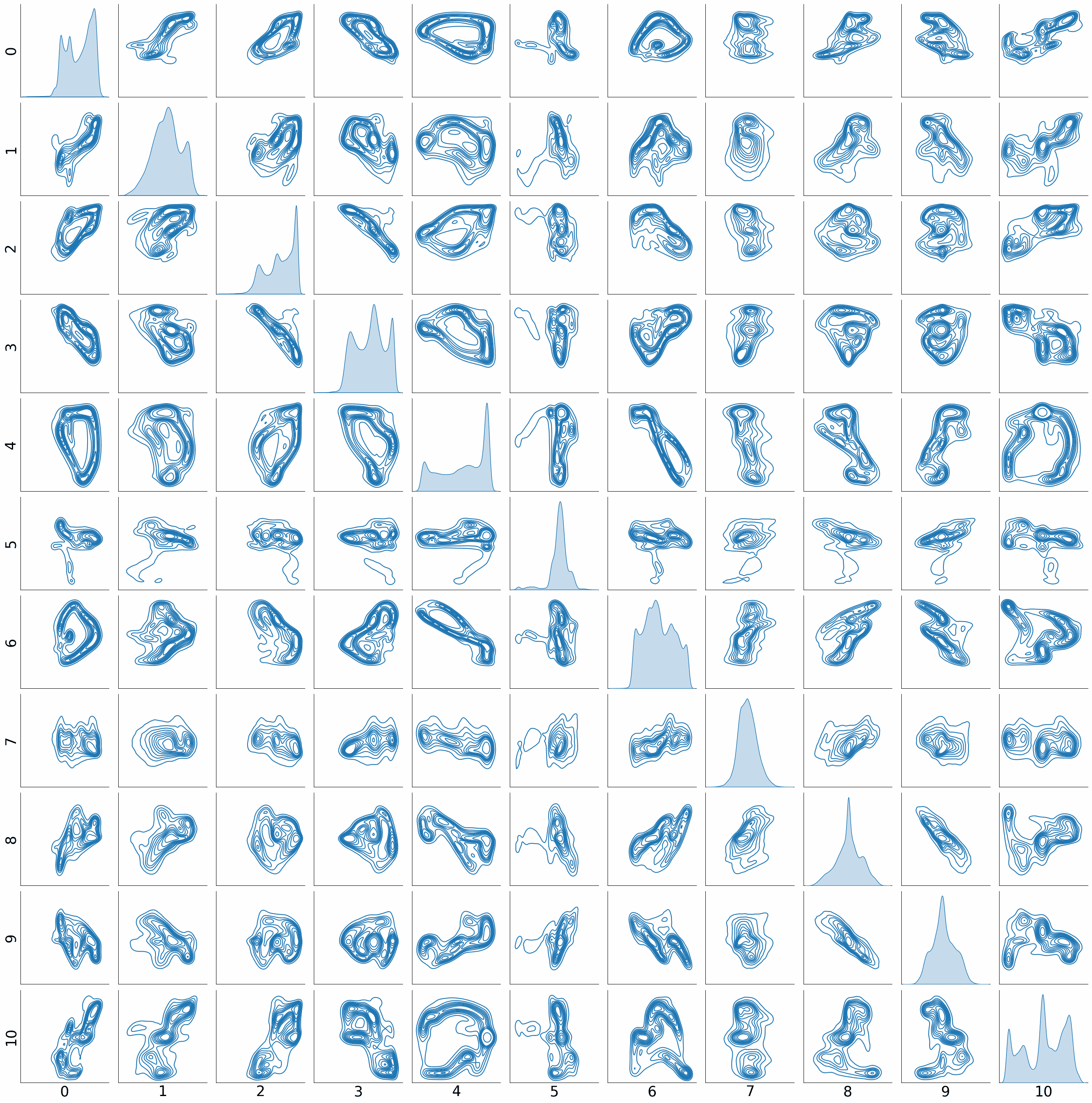}
    \caption{Distribution visualization using KDE for Hopper environment.}
    \label{fig:kde_hopper}
\end{figure}
\begin{figure}[ht]
    \centering
    \includegraphics[width=1\textwidth]{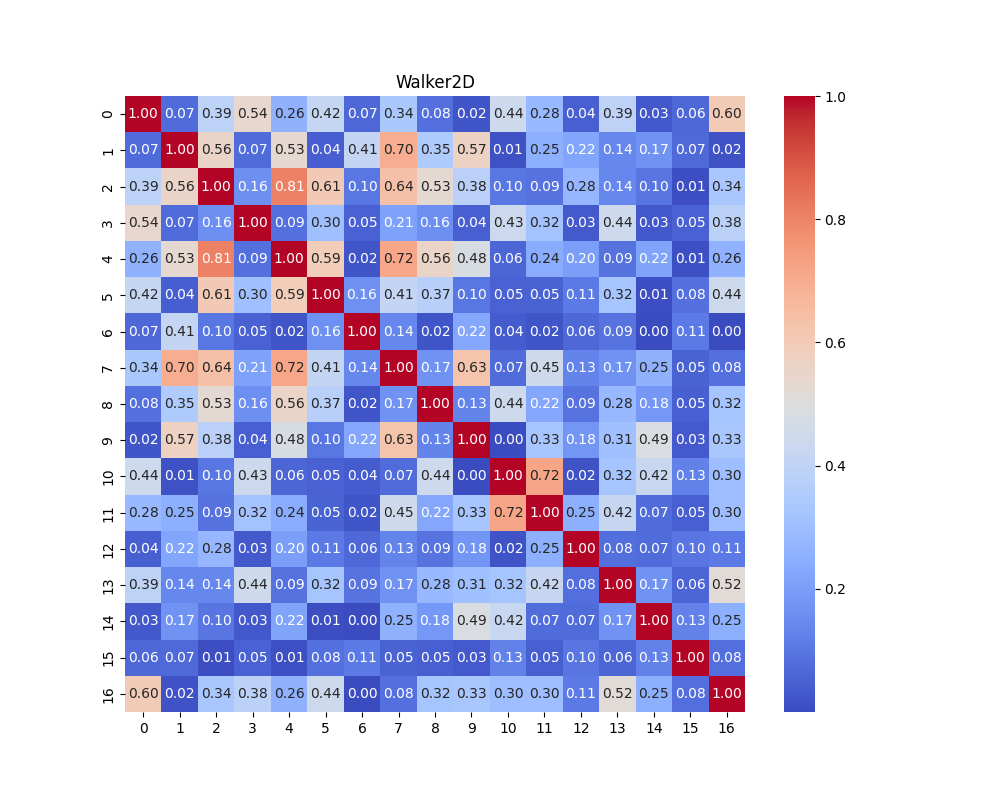}
    \caption{Correlation matrix of the states collected from Walker2D environment.}
    \label{fig:cm_walker2d}
\end{figure}
\begin{figure}[ht]
    \centering
    \includegraphics[width=1\textwidth]{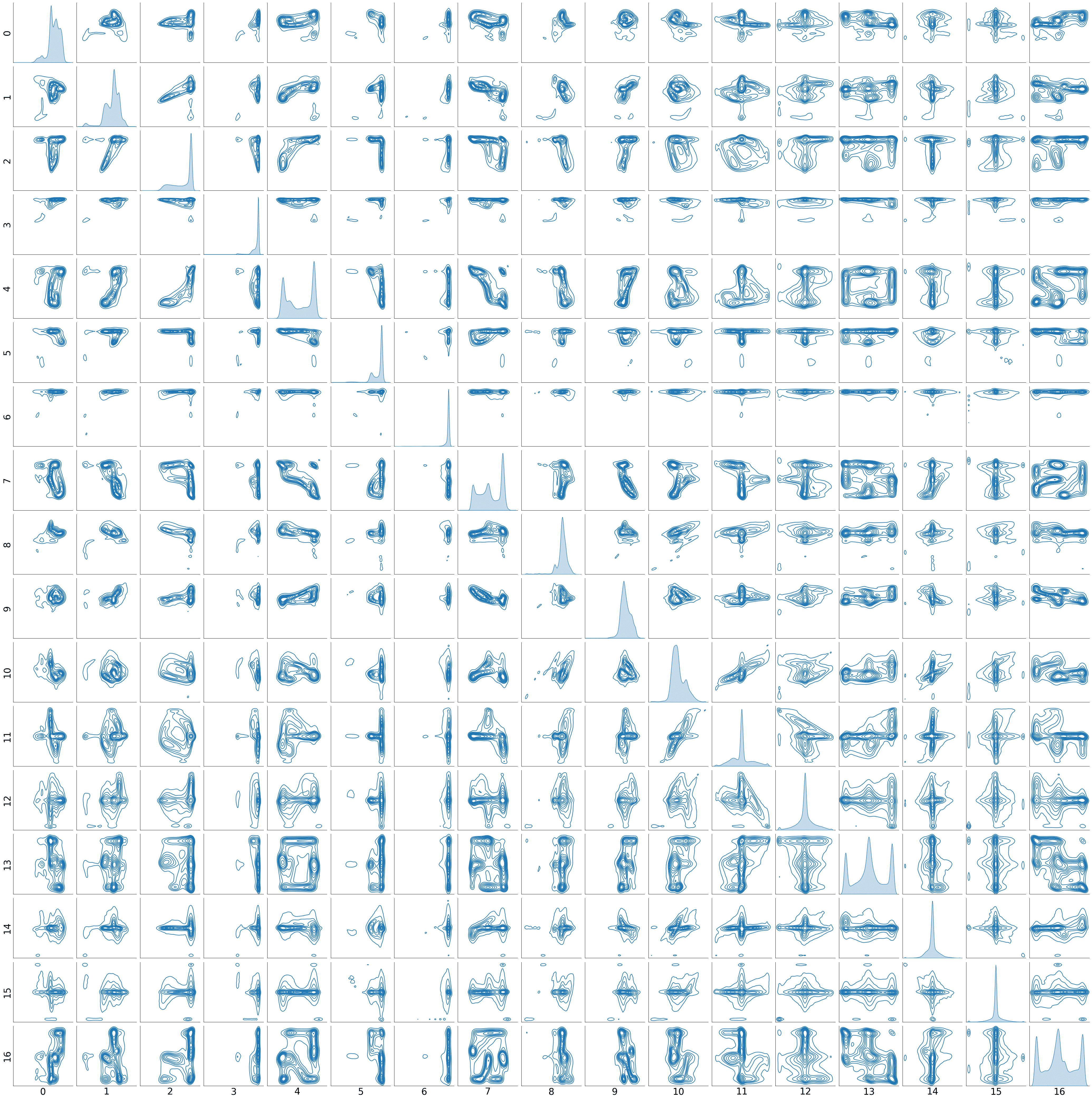}
    \caption{Distribution visualization using KDE for Walker2D environment.}
    \label{fig:kde_walker2d}
\end{figure}
\begin{figure}[ht]
    \centering
    \includegraphics[width=1\textwidth]{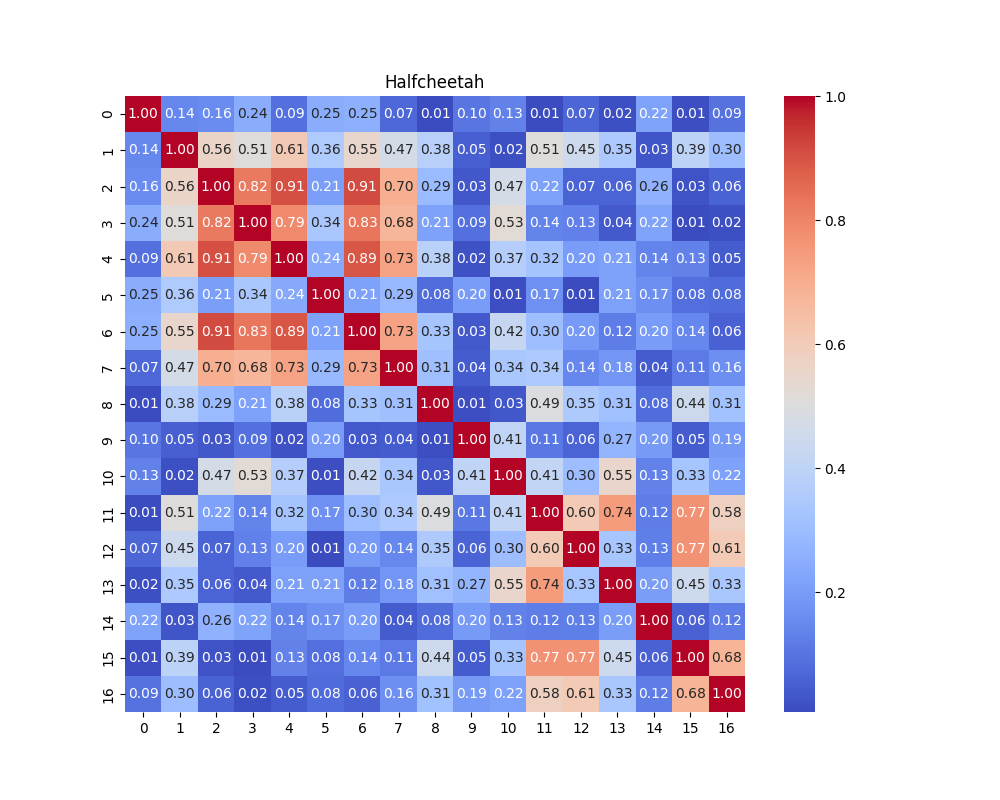}
    \caption{Correlation matrix of the states collected from HalfCheetah environment.}
    \label{fig:cm_halfcheetah}
\end{figure}
\begin{figure}[ht]
    \centering
    \includegraphics[width=1\textwidth]{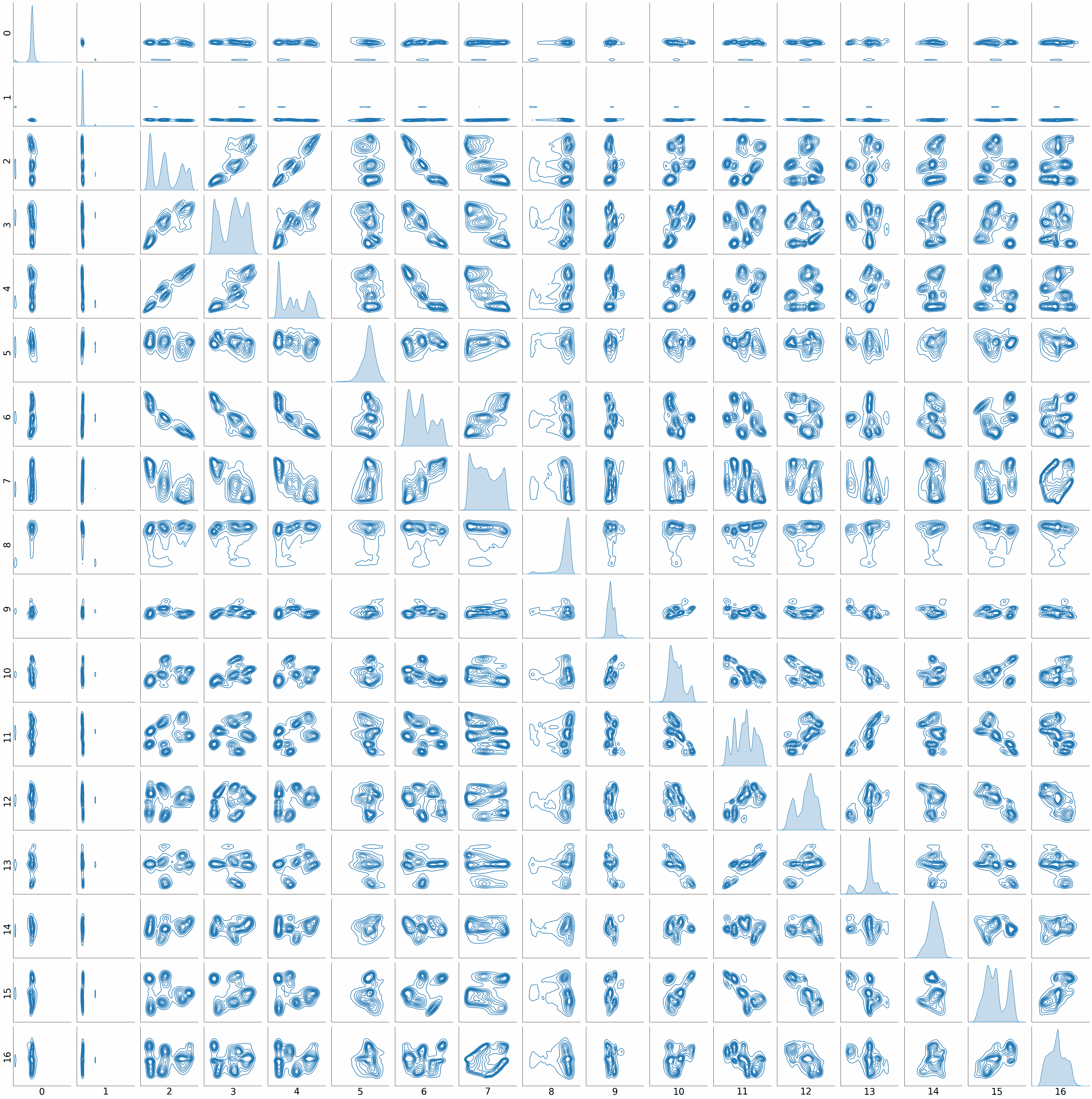}
    \caption{Distribution visualization using KDE for HalfCheetah environment.}
    \label{fig:kde_halfcheetah}
\end{figure}

In this section, we present a correlation matrix and the distribution shape of the real states, derived from 100k states collected during interactions between the victim policy and the Mujoco environment. We utilize Spearman's rank correlation matrix to analyze the relationships among different variables, as illustrated in \Cref{fig:cm_hopper}, \Cref{fig:cm_walker2d}, and \Cref{fig:cm_halfcheetah}. Additionally, we employ \ac{KDE}, a non-parametric approach, to estimate the probability density functions of these variables, shown in \Cref{fig:kde_hopper}, \Cref{fig:kde_walker2d}, and \Cref{fig:kde_halfcheetah}. We observe that the states are highly correlated and cannot be adequately described by a diagonal Gaussian distribution alone. This finding supports the capability of \ac{SI} to handle more complex input distributions beyond the scope of a diagonal Gaussian.

